\documentclass[a4paper,16pt]{article}
\usepackage{authblk} % need it for affiliation
\usepackage[english]{babel}
\usepackage[utf8x]{inputenc}
\usepackage[T1]{fontenc}
\usepackage{algorithm}
\usepackage[noend]{algpseudocode}
\usepackage{url}
\usepackage{xspace}
\usepackage{multirow}
\usepackage[a4paper,top=2cm,bottom=2cm,left=2cm,right=2cm,marginparwidth=1.75cm]{geometry}

\usepackage{amsmath}
\usepackage{graphicx}
\usepackage[colorinlistoftodos]{todonotes}

\usepackage{algorithm}
\usepackage{algpseudocode}

\usepackage{bm}
\usepackage{amssymb}
\usepackage{amsfonts}
\usepackage{pgfplots} \usetikzlibrary{plotmarks}
\usetikzlibrary{calc,backgrounds}

\DeclareMathAlphabet{\mathitbf}{OML}{cmm}{b}{it}
\def\jmath{j}

\def\jmath{j}

\usepackage[labelformat=simple]{subcaption}

\newcommand{\vTheta}{\varTheta}

\newcommand{\EXP}[1]{\mathbb{E}\left[#1\right ]}
\newcommand{\EXPt}[1]{\mathbb{E}_t\left[#1\right ]}
\newcommand{\Var}[1]{\mathbb{V}\left[#1\right ]}

\def\min{\mbox{min}}

\def\sol{c}

\def\D{\mathcal{D}}

\newcommand{\var}[1]{{\ensuremath{\mathrm{Var}}\mspace{-2mu}\left[#1\right]}}

\newcommand{\MSE}{\text{MSE}}

\def\xib{\bm{\xi}}

\def\bx{\mathbf{x}}

\newcommand{\conc}{c} % mass fraction
\newcommand{\pres}{p} % hydrostatic pressure
\newcommand{\poro}{\phi} % porosity
\newcommand{\perm}{{\mathbf{K}}} % permeability
\newcommand{\dens}{\rho} % density
\newcommand{\visc}{\mu} % viscosity
\newcommand{\dvel}{{\mathbf{q}}} % Darcy velocity
\newcommand{\grav}{{\mathbf{g}}} % gravity
\newcommand{\disp}{{\mathbf{D}}} % diffusion-disperion
\newcounter{theorem}
\newtheorem{remark}[theorem]{Remark}
\newtheorem{defn}[theorem]{Definition}
\newtheorem{theorem}{Theorem}

\newcommand{\Tau}{\mathcal{T}}

\title{Uncertainty quantification in the Henry problem using the multilevel Monte Carlo method}
\author[1]{Dmitry Logashenko}
\author[2]{Alexander Litvinenko}
\author[1,2]{Raul Tempone}
\author[3]{Ekaterina Vasilyeva}
\author[2]{Gabriel Wittum}
\affil[1]{KAUST, Thuwal-Jeddah, Saudi Arabia}
\affil[2]{RWTH Aachen, Aachen, Germany}
\affil[3]{Goethe-Universit\"at Frankfurt am Main, Germany}
\affil[ ]{\textit{litvinenko@uq.rwth-aachen.de,ekaterina.vasilyeva@gcsc.uni-frankfurt.de,
\{raul.tempone,dmitry.logashenko,gabriel.wittum\}@kaust.edu.sa}}
\makeatletter
\def\BState{\State\hskip-\ALG@thistlm}
\makeatother

\begin{document}
\maketitle

\footnotetext[2]{Corresponding author: litvinenko@uq.rwth-aachen.de}, %\email{alexander.litvinenko@kaust.edu.sa}. King Abdullah University of Science and %Technology (KAUST), Thuwal, Saudi Arabia.}

\begin{abstract}
{\color{black}
We investigate the applicability of the well-known multilevel Monte Carlo (MLMC) method to the class of density-driven flow problems, in particular the problem of salinisation of coastal aquifers. As a test case, we solve the uncertain Henry saltwater intrusion problem. Unknown porosity, permeability and recharge parameters are modelled by using random fields.
The classical deterministic Henry problem is non-linear and time-dependent, and can easily take several hours of computing time. Uncertain settings require the solution of multiple realisations of the deterministic problem, and the total computational cost increases drastically. Instead of computing of hundreds random realisations, typically the mean value and the variance are computed. The standard methods such as the Monte Carlo or surrogate-based methods is a good choice, but they compute all stochastic realisations on the same, often, very fine mesh. They also do not balance the stochastic and discretisation errors. These facts motivated us to apply the MLMC method. We demonstrate that by solving the Henry problem on multi-level spatial and temporal meshes, the MLMC method reduces the overall computational and storage costs.
To reduce the computing cost further, parallelization is performed in both physical and stochastic spaces. To solve each deterministic scenario, we run the parallel multigrid solver ug4 in a black-box fashion.}

%We use the solution obtained from the quasi-Monte Carlo method as a reference solution.
\end{abstract}

%\begin{keywords}
%\textcolor{black}{Corrections for the first reviewer are highlighted in RED.}\\
%\textcolor{black}{Corrections for the second reviewer are highlighted in BLUE.}\\
%\textcolor{black}{Corrections required by both reviewers are highlighted in VIOLET.}

%\end{keywords}

%\begin{AMS}
%15A69,  %       Multilinear algebra, tensor products
%65F10,  % Iterative methods for linear systems
%60H15,  %  Stochastic partial differential equations
%60H35,  %  Computational methods for stochastic equations 
%65C30  %  Stochastic differential and integral equations
%\end{AMS}

%

\section{Introduction}
\label{sec:1}

{\color{black} We investigate the applicability of the well-known multilevel Monte Carlo (MLMC) method to density-driven flow problems, in particular the problem of salinisation of coastal aquifers. As a numerical test case, we consider the well-known Henry problem in a stochastic setting. The novelty here is that parameters such as porosity, permeability and recharge are not deterministic but stochastic (uncertain). The reason for the presence of uncertainties is the lack of knowledge, inaccurate measurements,  and the inability to measure parameters at each spatial or temporal location. Although the deterministic Henry problem is well-known, the difficulties here are that it is not clear how input uncertainties propagate through the non-linear, time-dependent problem. The solution to be found is then the mean value and the variance of the salt mass fraction, both are functions of space and time. 
An accurate estimate of the output uncertainties can facilitate a better understanding of the problem, better decisions, and improved control and design of the experiment.
From an implementation point of view, we test how easy it is to couple multigrid and multilevel Monte Carlo solvers.

\textbf{Salinisation of coastal aquifers.} The intrusion of salt water occurs when the sea level rises and the salt water moves onto the land \cite{SWLRHEGW-SaltwaterInNorthSea2018}. This usually happens during storms, floods, droughts or when saltwater intrudes into freshwater aquifers and raises the water table. As groundwater is an essential resource for food and irrigation, its salinisation can be disastrous. Many hectares of farmland could be lost because it becomes too wet or too salty to grow crops. Therefore, accurate modelling of different saline flow scenarios is essential \cite{Abarca07,SWLRHEGW-SaltwaterInNorthSea2018} to help farmers and researchers develop strategies to improve soil quality and reduce the impact of saltwater intrusion.

\textbf{Modeling.} The saltwater flow is density driven and described by a system of time-dependent nonlinear partial differential equations (PDEs). It is characterised by convection dominance and can exhibit very complicated behaviour \cite{Voss_Souza}. Uncertain parameters can have a strong influence on the flow and transport of salt. Random fields are used to model these parameters. 

There are a number of studies where authors have modelled uncertainties in reservoirs (see \cite{OverviewUncert93,SoilOverview16}). The link between stochastic methods and hydrogeological applications was made in \cite{NowakStochMethods18}, where the authors analysed a collaboration between academics and water suppliers in Germany and made recommendations for optimisation and risk assessment. The basics of stochastic hydrogeology and an overview of stochastic tools and uncertainty management are described in \cite{rubin2003applHydro}. 

The review \cite{TARTAKOVSKYI_Risk13} deals with hydrogeological applications of recent advances in uncertainty quantification, probabilistic risk assessment and decision making under uncertainty. The author reviews probabilistic risk assessment methods in hydrogeology under parametric, geological and model uncertainties. Density-driven vertical transport of saltwater through the freshwater lens on the island of Baltrum (Germany) is modelled in 
\cite{POST17_Density-driven}.

In \cite{Laattoe2013_SeawaterIntr}, the authors examined the implications of transgression for a range of seawater intrusion scenarios based on simplified coastal freshwater aquifer settings. They stated that vertical intrusion during transgressions could involve density-driven convective processes, causing substantially greater amounts of seawater to enter the aquifer and create more extensive intrusion than horizontal seawater intrusion in the absence of transgression.

\textbf{History and preliminary results.} This work is an extension and continuation of our preliminary works \cite{Litvinenko_Gamm23,Litv_Uncecomp23}. The original Henry saltwater intrusion problem was introduced by H.R. Henry in the 1960s (see \cite{henry1964effects}) and became a benchmark for numerical solvers for groundwater flow (see \cite{Voss_Souza,Simpson04_Henry,Simpson2003,Dhal_review22}. In \cite{Riva2015}, the authors use the generalised polynomial chaos expansion approximation to investigate how incomplete knowledge of system properties affects the assessment of global quantities. In particular, they estimated the propagation of input uncertainties into a few dimensionless scalar parameters.

Hydrogeological formations typically have complex and heterogeneous structures.
These formations may consist of several layers of porous media with different porosity and permeability coefficients
(cf.\ \cite{ReiterLogashenkoVogelWittum2017, ScheiderKroehnPueschel2012}).
Measurements of the layer positions and their thicknesses are only possible with a certain degree of error, and
average parameters are typically assumed for the materials within the layers. These layers are therefore excellent candidates for random field modelling. Furthermore, due to the non-linearities in the problem, computing with the averaged parameters does not necessarily lead to the mathematical expectation of the solution that can be obtained by averaging over all scenarios.

\textbf{Methods.} Many techniques can be used to quantify uncertainties. One classical method is Monte Carlo (MC) sampling. Although it is dimension independent, it converges very slowly, with a convergence rate of $\mathcal{O}(\frac{1}{\sqrt{N}})$. This method may not be affordable for time-consuming simulations. However, even modern techniques such as surrogate models and stochastic collocation require a few hundred to a few thousand time-consuming simulations and assume a certain smoothness of the quantity of interest (QoI).

Another class of methods is the class of perturbation methods \cite{CREMER15_Fingers}. The idea is to decompose the QoI with respect to (w.r.t.) random parameters in a Taylor series. The higher order terms can be neglected for small perturbations, simplifying the analysis and numerical tests. These methods assume that the random perturbations are small. For larger perturbations, these methods usually do not work.

Other methods to compute the desired statistics of the QoI are direct integration methods, such as the quasi-MC (QMC) \cite{caflisch98}, collocation methods and surrogate-based (generalized polynomial chaos approximation \cite{xiuKarniadakis02a,Litvinenko-UQ-2021, LitLog3D_20, sudret2008global, Sudret2021Review_PCE}
and stochastic Galerkin \cite{babuska2004galerkin,GiraldiLitv14,wahnert-stochgalerkin-2014}) methods. Direct methods compute statistics directly by sampling uncertain input coefficients and solving the corresponding PDEs, whereas the surrogate-based method computes a cheap functional (polynomial, exponential, or trigonometrical) approximation of the QoI. Examples of the surrogate-based methods are radial basis functions \cite{liu2014,bompard2010,Loeven2007,giunta2004}, sparse polynomials \cite{
chkifa-adapt-stochfem-2015,Sudret_sparsePCE,DolgLitv15}, and polynomial chaos expansion \cite{Habib09_PCE, ConradMarzouk13, Dongbin}. 
Surrogate methods are using some well-known functions such as the multivariate Legendre, Hermite, Chebyshev or Laguerre functions \cite{OLADYSHKIN_PCE,xiuKarniadakis02a}. Advantages of surrogate methods are: Once the model is constructed, it is easy to sample, and all the samples are almost free (much cheaper than sampling the original stochastic PDE). The non-trivial part of surrogate models is to define which polynomial order is needed and how accurately all coefficients should be computed. Another difficulty is that not every function can be well approximated by a polynomial.

Sparse grid methods to integrate high-dimensional integrals are considered in \cite{smoljak63, Griebel_Bungartz, Griebel, spiterp, novakRitter97, gerstnerGriebel98-numint, novakRitter99-simple, ConradMarzouk13,petrasSmolpak}. An idea to generate goal-oriented adaptive spatial grids and use them in the multilevel MC (MLMC) framework was presented in \cite{EIGEL14,BECK22}.

The quantification of uncertainties in stochastic PDEs could be a significant challenge due to a) the large possible number of involved random variables and b) the high cost of each deterministic solution of the governed PDE. The MC quadrature and its variance-reduced variants have a dimension-independent error convergence rate $\mathcal{O}(N^{-\frac{1}{2}})$, and the QMC has the worst-case rate $\mathcal{O}(\log^M(N)N^{-1})$, where $N$ is the number of samples, and $M$ indicates the dimension of the stochastic space \cite{matthies2007}. The MC method is not affected by the dimension of the
integration domain, such as collocations on sparse or full grid methods \cite{babuska_collocation, nobile-sg-mc-2015}. A numerical comparison of other QMC sequences is presented in \cite{radovic1996}.

%The family of MLMC methods does not have such limitations.

This work has the following structure. Section~\ref{sec:Model} describes the Henry problem and numerical methods to solve it. The well-known MLMC method is reviewed in Section~\ref{sec:MLMC}. Next, Section~\ref{sec:numerics} details the numerical results, which include the numerical analysis of the Henry problem, the computation of different statistics, the performance of the MLMC method, and the practical performance of the parallel ug4 solver for the Henry problem \cite{henry1964effects,Simpson04_Henry} with uncertain coefficients. Finally, we conclude this work with a discussion in section~\ref{sec:Conclusion}.

\textbf{Our contribution and main results:} We coupled the MLMC method with the multigrid method to estimate the propagation of uncertainties in the Henry problem. We used random fields to model unknown porosity, permeability and recharge. In our model, porosity and permeability are spatially dependent, random, multi-scale and have two layers. The recharge is time dependent and uncertain. We have not simplified our equations with the Boussinesq approximation, i.e. we have considered a more general case. We used MLMC to compute the mean and variance of the mass fraction. In addition, several other quantities of interest were computed, such as the total freshwater integral, the salt integral, the mass fraction computed at a point and integrated over a small subdomain. We investigated which of these QoIs were more appropriate to be computed by the MLMC method. To further speed up the computational process, we run all the MLMC random simulations and each individual simulation in parallel.}\
This work is not only of theoretical interest, but also of practical interest; our modelling and solution allow us to answer the following questions:
\begin{enumerate}
\item How long can a particular drinking water well be used (i.e. when will the mass fraction of salt exceed a critical threshold)?
\item Which part of the aquifer has particularly high uncertainty?
\item What is the probability that the salt concentration at a given location and time will exceed a threshold value?
\item What is the mean scenario (and its variations)?
\item What are the extreme scenarios for the concentration of the salt?
\item How do the uncertainties change with time?
\end{enumerate}
 
%%%%%%%%%%%%%%%%%%%%%%%%%%%%%%
To the best of our knowledge, we are not aware of any other studies where Henry's problem \cite{henry1964effects,Simpson04_Henry} has been solved in parallel using MLMC methods with uncertain porosity, permeability and recharge parameters.
\section{Henry Problem with Uncertain Parameters}
\label{sec:Model}
\subsection{Problem setting}
\label{subsec:Henry}
In coastal aquifers, saline seawater intruding into the formation from one side (the sea side) displaces pure water from the other side due to water recharge from land sources and precipitation. Because of its higher density, seawater tends to infiltrate along the bottom of the aquifer. This process can reach a steady state, but may be time-dependent due to the periodicity of the recharge or the control of the pumping rate from the wells. Accurate simulation of salinisation is essential for predicting the availability of water resources. However, the accuracy of such predictions strongly depends on the hydrogeological parameters of the formation and the geometry of the computational domain, denoted by $\D$.

% {\color{black} For some alternative formulation, namely, the hydraulic-head formulation, see \cite{Langevin20}.}

The aquifer $\D \subset \mathbb{R}^d$, $d \in \{2, 3 \}$, can be modelled as an immobile porous matrix filled with liquid phase --- a solution of salt in water. Due to the inhomogeneous density distribution, gravity induces the motion of the liquid phase. This motion transports the salt, which would otherwise be subject to molecular diffusion.

\begin{figure}[t!]
\begin{center}
  \includegraphics[width=0.5\textwidth]{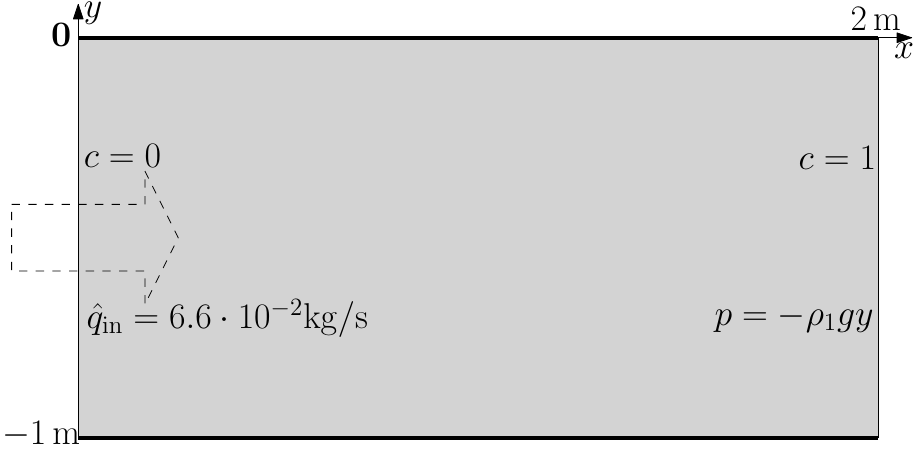}\;
   \includegraphics[width=0.45\textwidth]{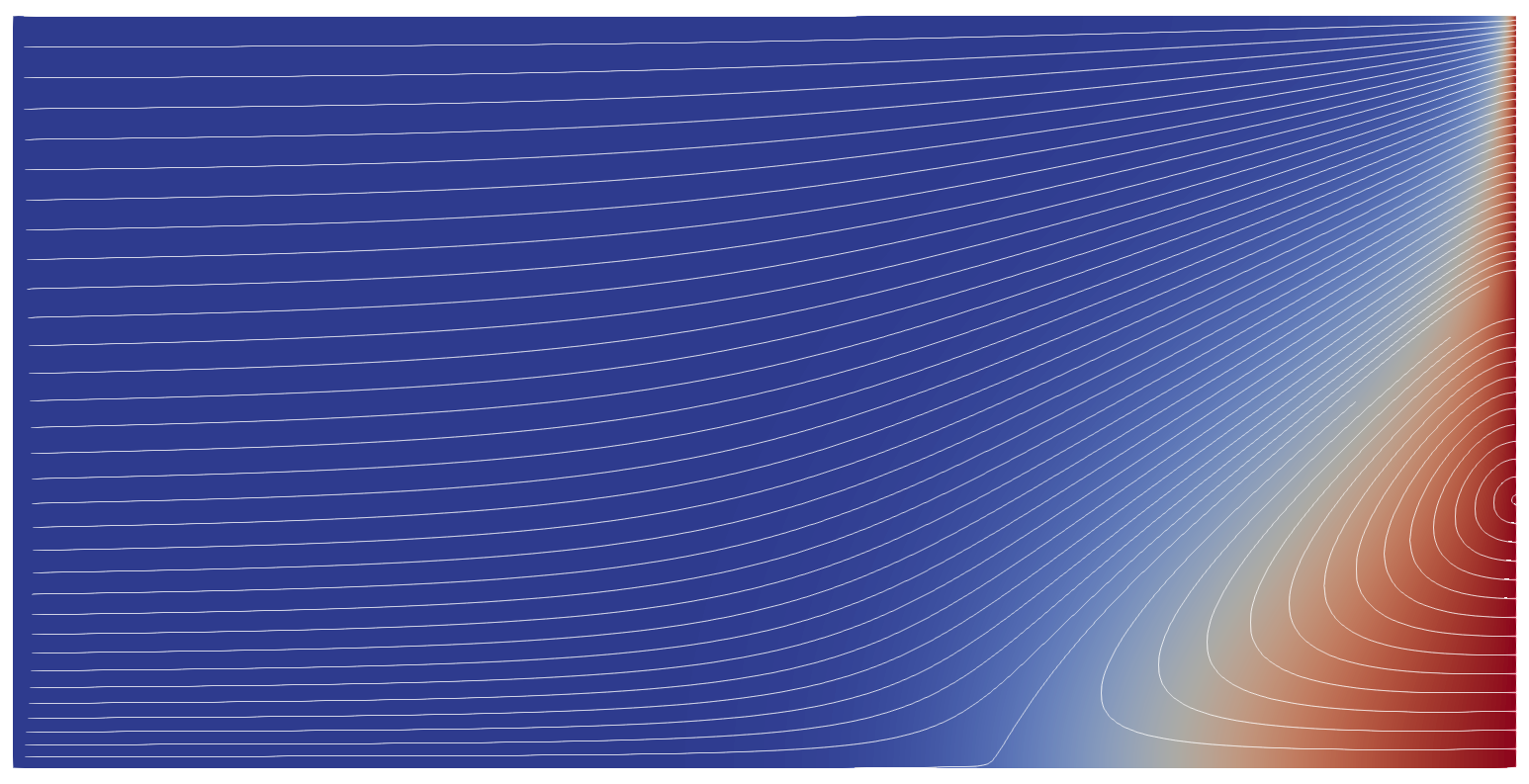}
  \caption{(left) Computational domain $\mathcal{D}:=[0,2]\times [-1,0]$; (right) the mass fraction $\sol\in [0,1]$ and the streamlines of the velocity field $\dvel$ for the undisturbed Henry problem at $t = 6016$ $\mathrm{s}$. {\color{black} The dark red colour corresponds to $\sol=1$ and the blue colour to $\sol=0$.}}
    % One realization of $\conc(t,\bx)$ at $t = 6016$ s. and 15 pre-selected points with small subdomains around them. The size of the subdomain around point $(x,y)$ is $[x-0.1,x+0.1]\times [y-0.1,y+0.1]$.
    % % (Right) One realization of the mass fraction $\conc(t,\bx)$ and the streamlines of the velocity field $\dvel$ for the undisturbed Henry problem at $t = 6016$ $\mathrm{s}$.}
    \label{fig:Henry2d-scheme}
\end{center}    
\end{figure}

A simple but very illustrative model of coastal aquifers is the so-called Henry problem, first considered in \cite{henry1964effects}. In this two-dimensional setting, the aquifer is represented by a rectangular domain $\D = [0, 2] \times [-1, 0]$ $[\mathrm{m}^2]$ completely saturated with the liquid phase (Fig.~\ref{fig:Henry2d-scheme}). The salty seawater intrudes from the right and the pure water recharges from the left. The top and bottom are assumed to be impermeable. Analogous settings with partially saturated domains are considered in \cite{Stoeckl}.

The mass conservation laws for the whole liquid phase and the salt give the following equations
\begin{eqnarray}
 \label {e_cont_eq}
 \partial_t (\poro \dens) & + & \nabla \cdot (\dens \dvel) = 0, \\
 \label {e_tran_eq}
 \partial_t (\poro \dens \conc) & + & \nabla \cdot (\dens \conc \dvel - \dens \disp \nabla \conc) = 0,
\end{eqnarray}
where $\poro: \D \to \mathbb{R}$ denotes the porosity, $\perm: \D \to \mathbb{R}^{d \times d}$ represents the permeability of the porous matrix, $\conc (t, \mathbf{x}): [0, +\infty) \times \D \to [0, 1]$ is the mass fraction of the salt (or of the brine) in the solution, $\dens = \dens (\conc)$ indicates the density of the liquid phase, and $\disp (t, \mathbf{x}, \dvel): [0, +\infty) \times \D {\color{black} \times \mathbb{R}^d } \to \mathbb{R}^{d \times d}$ denotes the molecular diffusion and mechanical dispersion tensor. For the velocity $\dvel (t, \mathbf{x}): [0, +\infty) \times \D \to \mathbb{R}^d$, we assume Darcy's law:
\begin{eqnarray} \label {e_Darcy_vel}
 \dvel = - \frac{\perm}{\visc} (\nabla \pres - \dens \grav),
\end{eqnarray}
where $\pres = \pres (t, \mathbf{x}): [0, +\infty) \times \D \to \mathbb{R}$ is the hydrostatic pressure, $\visc = \visc (\conc)$ denotes the viscosity of the liquid phase, and $\grav = (0, \dots, 0, - 9.8)^T \in \mathbb{R}^d$ represents the gravity vector. Inserting (\ref {e_Darcy_vel}) into (\ref {e_cont_eq}--\ref {e_tran_eq}) results in a system of two time-dependent PDEs in the unknowns $\conc$ and $\pres$. This system should be closed with boundary conditions for $\conc$ and $\pres$ and an initial condition for $\conc$.

\begin{table}[b]
\begin{center}

 \begin{tabular}{|l|l|l|} \hline
  Parameter & Values and Units & Description \\ \hline
  $\hat{\phi} := \EXP{\phi}$ & 0.35 [-] & mean value of porosity \\ \hline
  $D$ & $18.8571\cdot 10^{-6}$ [$\mathrm{m}^2 \cdot \mathrm{s}^{-1}$] & diffusion coefficient in the medium \\ \hline
  $\hat{K}$ & $1.020408\cdot 10^{-9}$ [$\mathrm{m}^2$] & reference value of permeability \\ \hline
%  $g$ & $9.8$ [$\mathrm{m} \cdot \mathrm{s}^{-2}$] & gravity \\ \hline
  $\rho_0$ & $1000$ [$\mathrm{kg} \cdot \mathrm{m}^{-3}$] & density of pure water \\ \hline
  $\rho_1$ & $1024.99$ [$\mathrm{kg} \cdot \mathrm{m}^{-3}$] & density of brine \\ \hline
  $\mu$ & $10^{-3}$ [$\mathrm{kg} \cdot \mathrm{m}^{-1} \cdot \mathrm{s}^{-1}$] & viscosity \\ \hline
  $\kappa_{KC}$ & $2.088415 \cdot 10^{-8}$  [$\mathrm{m}^2$] & scaling in the Kozeny-Carman \\ \hline
 \end{tabular}
 \caption{Parameters of the considered density-driven flow problem}
 \label{tab:HenryParam}
\end{center}
\end{table}

Following the classical setting in {\color{black} \cite{Voss_Souza}}, for this variant of the Henry problem {\color{black} (see also \cite{Simpson2003,Simpson04_Henry})}, we set
\begin {eqnarray} \label {e_lin_density}
 \dens (\conc) = \dens_0 + (\dens_1 - \dens_0) \conc, \qquad  \qquad \visc = \text{const}
\end {eqnarray}
and
\begin {eqnarray} \label {e_mol_diff}
 \disp = \poro D \mathbf{I}, \; \text{with a scalar}\;\; D \in \mathbb{R}, \; \text{and}\;\;\mathbf{I}\in\mathbb{R}^{d\times d}\; \text{the identity matrix.}
\end {eqnarray}
Furthermore, we assume the isotropic permeability
\begin {eqnarray*} %\label {e_perm_of_poro}
 \perm = K \mathbf{I}, \qquad K \in \mathbb{R}.
\end {eqnarray*}
This setting is consistent with the problem setting in \cite{Voss_Souza}. However, we do not assume the Boussinesq approximation and keep the density variable for all terms.

For the initial conditions, we set
\begin {equation}
 \left. \conc \right |_{t = 0} = 0.
\end {equation}

The boundary conditions are presented in Fig.~\ref {fig:Henry2d-scheme} (left). On the right side of the domain, we impose Dirichlet conditions for $\conc$ and $\pres$ variables that represent the adjacent seawater aquifer:
\begin {equation}
	\left. \conc \right |_{x=2} = 1, \qquad \left . \pres \right |_{x=2} = - \rho_1 g y.
\end {equation}
On the left side, we prescribe the inflow of fresh water:
\begin {equation}
	\left. \conc \right |_{x=0} = 0, \qquad \left . \dens \dvel \cdot \mathbf{e}_x \right |_{x=0} = \hat{q}_{\mathrm{in}},
\end {equation}
where $\mathbf{e}_x = (1, 0)^\top$, and $\hat{q}_{\mathrm{in}}$ is a constant. For the classical formulation of the Henry problem, this value was set to $\hat{q}_{\mathrm{in}} = 6.6 \cdot 10^{-2}$ $\mathrm{kg}/\mathrm{s}$ in \cite{Voss_Souza} or $\hat{q}_{\mathrm{in}} = 3.3 \cdot 10^{-2}$ $\mathrm{kg}/\mathrm{s}$ in \cite{Simpson04_Henry,Simpson2003}. The Neumann zero boundary conditions are imposed on the upper and lower sides of $\mathcal{D}$.

\begin{figure}[t!]
\begin{center}
 \includegraphics[width=0.65\textwidth]{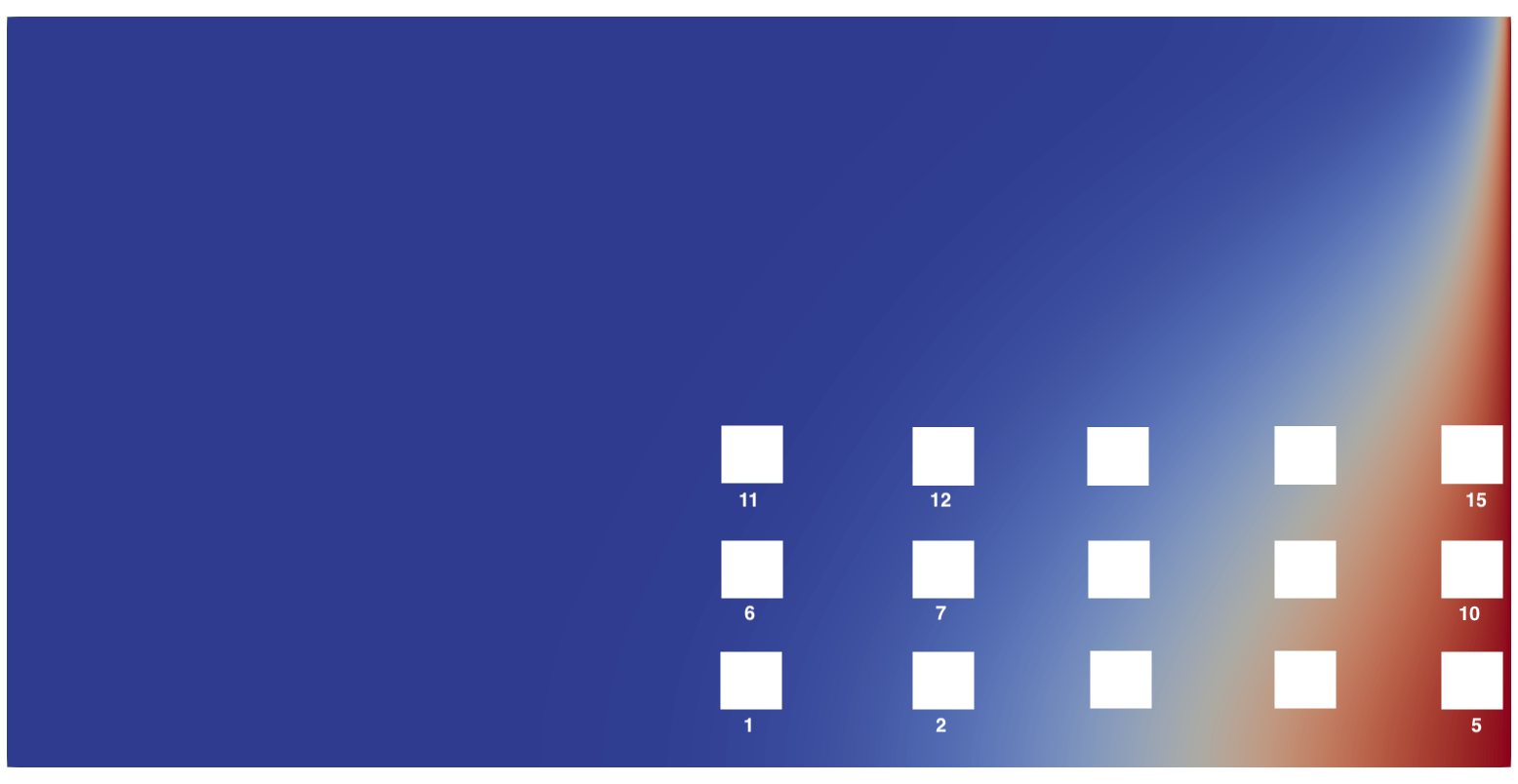}
  \caption{Positions of the 15 pre-selected points with small subdomains around them. The size of the subdomain around point $(x_i,y_i)$ is $[x_i-0.1,x_i+0.1]\times [y_i-0.1,y_i+0.1]$.}
    \label{fig:Henry2d-scheme_points}
\end{center}    
\end{figure}

An example of $\conc(t,\bx)$ and the flow for the parameters from Table~\ref {tab:HenryParam} (i.e. $\phi = \hat{\phi}$ and $K = \hat{K}$) is shown in Fig.~\ref {fig:Henry2d-scheme}(right). The dark red colour corresponds to $\conc=1$ (salt water) and the dark blue colour corresponds to $\conc=0$ (pure water). Due to its higher density, the saltwater intrudes into the aquifer at the bottom right. It is pushed back by the lighter pure water coming from the left. This process creates a vortex in the flow at the bottom right of the domain. The salt water enters at the lower part of the right boundary and diverts upwards and to the right, back to the sea, forming a salt triangle. This flow does not transport the salt to the left part of the domain. The salt moves further to the left by diffusion and dispersion and is washed out by the recharge. In the classical formulation, this salt triangle initially grows with time but reaches a steady state (see \cite{Voss_Souza,Simpson04_Henry,Simpson2003}). However, the initial non-stationary phase can take considerable time. Investigation of this phase is particularly important to understand the behaviour of the system when the intensity of recharge varies, as may occur, for example, due to climate change. We consider integrals over the whole domain $\D$, which describes the total amount of pure water and the total amount of salt (as in equations~\eqref{eq:integral_fw}--\eqref{eq:integral_box}). We also compute local integrals over 15 small rectangular subdomains $\Delta_1,\ldots,\Delta_{15}$, $\Delta_i := [x_i-0.1,x_i+0.1]\times [y_i-0.1,y_i+0.1]$ around preselected points defined below. The list of selected points follows:
\begin{align}
\label{eq:12points} 
\{\bx_i := (x_i,y_i)_{i=1,\ldots,15}\}&=
\{
(0.90, -0.95),
	(1.15, -0.95),
	(1.40, -0.95),
	(1.65, -0.95),
	(1.90, -0.95),\\ \nonumber
	&
    (0.90, -0.75),
	(1.15, -0.75),
	(1.40, -0.75),
	(1.65, -0.75),
	(1.90, -0.75),\\
	 \nonumber 
	&
    (0.90, -0.50),
	(1.15, -0.50),
	(1.40, -0.50),
	(1.65, -0.50),
	(1.90, -0.50)
\}
 \end{align}
% %\{p_1,p_2,\ldots,p_{12}\}:
% \{(x,y)_{i=1,\ldots,12}\}&=\{(1.10, -0.95),
% 	(1.35, -0.95),
% 	(1.60, -0.95),
% 	(1.85, -0.95),
% 	(1.10, -0.75),
% 	(1.35, -0.75), \label{eq:12points}\\ 
% 	&(1.60, -0.75),
% 	(1.85, -0.75),
% 	(1.10, -0.50),
% 	(1.35, -0.50),
% 	(1.60, -0.50),
% 	(1.85, -0.50).\} \nonumber 
%One could also consider other points.
(see Fig.~\ref{fig:Henry2d-scheme_points}.) The reason we consider 15 points (and 15 small subdomains) is that not all points are ``interesting'', i.e., not all points have significant variation in $\sol$. MLMC reduces the variance, but if the initial variance is {\color{black} smaller according to the prescribed tolerance}, there is no point in using MLMC.
%We remind that the mass fraction $\conc$ at each point $\bx$ is a function of time. 

The knowledge of  $\conc$ at these spatial points may help track salinity changes over time in groundwater wells and understand which areas in the aquifer are most vulnerable.
% Farmers can use this information to take action, such as decreasing salinity or adapting strategies by planting salt-tolerant crops.

%!!!
%
%
\subsection{Modeling porosity, permeability, and recharge}
\label{subsec:PorosityVar}
The primary sources of uncertainty are the hydrogeological properties of the porous medium --- porosity ($\poro$) and permeability ($\perm$) fields of the solid phase --- and the freshwater recharge flux $\hat{q}_x$ through the left boundary. 
%Other uncertainties are not considered in this work.    
The QoIs are related to the mass fraction $\sol$, a function of $\poro$, $\perm$, and the recharge. We model the uncertain $\poro$ using a random field and assume $\perm$ to be isotropic and dependent on $\poro$:
\begin {eqnarray} \label {e_perm_of_poro}
 \perm = K \mathbf{I}, \qquad K = K (\poro) \in \mathbb{R}.
\end {eqnarray}
The distribution of $\poro(\bx,\xib)$, $\bx\in \D$, is determined by a set of stochastic parameters {\color{black} $\xib=(\xi_1,\ldots,\xi_M)$}. Each component $\xi_i$ is a random variable depending on a random event $\omega$. For concision, we skip $\omega$ and write $\xib:=\xib(\omega)$.

The dependence in \eqref{e_perm_of_poro} is specific for every material.
%and there is no a general law. 
We refer to \cite{Panda_Lake_Perm_vs_Por,Pape_Clauser_Iffland_1999,Costa_2006} for a detailed discussion. In the proposed model, we use a Kozeny--Carman-like dependence
\begin{eqnarray} 
\label{e_perm_Kozeny_Carman}
 K (\poro) = \kappa_{KC} \cdot \dfrac {\poro^3} {1 - \poro^2},
\end {eqnarray}
where $\kappa_{KC}$ is a constant scaling factor.
% \begin{remark}
% Typical porosity values, which we have found in the literature are: well sorted sand $0.25-0.5$, poorly sorted sand $0.15-0.3$, clay $0.4-0.6$, crystalline rock $0.001-0.01$, fractured rock $0.01-0.05$.
% \end{remark}
The inflow flux is kept constant across the left boundary but depends on the stochastic variable $q_{\mathrm{in}}$. We also assume that the inflow flux is independent of $\poro$ and $\perm$.

{\color{black}
\begin{remark}
Equation (\ref{e_mol_diff}) is a simplification typically adopted when the Henry problem is used as a benchmark, see \cite{Voss_Souza,Simpson2003,Simpson04_Henry}. It is then assumed that $D$ is not only the molecular diffusion coefficient but also includes some dispersivity. In a more accurate model, the dispersion tensor should depend on the Darcy velocity $\dvel$, for example using the Bear-Scheidegger law (see \cite{Scheidegger-Dispersion,Bear-2}). However, we note that the parameters of mechanical dispersion depend on the configuration of the porous matrix (e.g. grain size) and therefore cannot be considered independent of porosity and permeability in stochastic modelling. The influence of dispersion on the propagation of uncertainties is therefore a non-trivial issue even from a modelling point of view. However, it should be addressed in the future.
\end{remark}
}

% In the experiments \hat{q}_x = \hat{q}_x (\theta_3) = -6.6e-2 * (1 + 0.5 * \theta_3ls)
%A discretisation of these uncertain parameters see in Sect.~\ref{sec:Discret}.
%
%
%\subsection{Quantities of interest}
%\label{sec:QoI}
%$P(c>c^*)>\varepsilon$, where $c^*$ is some critical value, $0<\varepsilon<1$
%
%

\subsection{Numerical methods for the deterministic problem}
\label{sec:Num}
The system (\ref{e_cont_eq}--\ref{e_tran_eq}) is numerically solved in the domain $\D \times [0, T]$, where $\times$ denotes the Cartesian product. $\D$ is covered with a grid $\D_h$ of quadrilaterals with mesh size $h$. {\color{black} Equations (\ref{e_cont_eq}--\ref{e_tran_eq}) are discretized on $\D_h$ using a vertex-centered finite-volume scheme as presented in \cite{Frolkovic-DeSchepper-ConvDom,Frolkovic-ConsVel,Frolkovic-Knaber-ConsVel}.} The number of degrees of freedom associated with $\D_h$ is denoted by $n$. There are two degrees of freedom per grid vertex in $\D_h$: one for the mass fraction $\sol$ and another for the pressure $\pres$. Note that $h = \mathcal{O} (n^{-1/d})$, $d=2$. We use the implicit Euler method with a time step $\tau$ for time discretization. The number of the computed time steps is $r = T / \tau$.
% with a fixed time step length to simplify evaluations of the stochastic quantities.
Solution of the discretized system yields the approximation $\sol_{h,\tau}$ for $\sol$ on this grid.

We use partial upwind for the convective terms (cf.\ \cite{Frolkovic-DeSchepper-ConvDom}). 
Therefore, the discretization error is of the second order w.r.t. the spatial mesh size $h$. However, Equation \eqref{e_tran_eq} is essentially convection dominated. For the grids in the numerical experiments, the observed reduction of the discretization error after grid refinement corresponds to the first order. Furthermore, the Euler method provides the first-order discretization error w.r.t. $\tau$. Thus, we assume the first-order dependence of the discretization error w.r.t. $h$ and $\tau$, i.e., as $d = 2$,
\begin {equation} \label {eq:1st_ord_conv}
 \| \conc - \conc_{h,\tau} \|_2 = \mathcal{O}(h + \tau) = \mathcal{O} (n^{-1/2} + r^{-1}),
\end {equation}
which is consistent with our numerical experiments.

The implicit time-stepping scheme provides unconditional stability but requires the solution of an extensive nonlinear algebraic system of the discretized equations with $n$ unknowns in every time step. The Newton method is used to solve this system. Linear systems inside the Newton iteration are solved using the BiCGStab method (cf.\ \cite {Templates}) preconditioned with the geometric multigrid method (V-cycle, cf.\ \cite{Hackbusch85}). In the multigrid cycle, the ILU${}_\beta$-smoothers \cite{Hackbusch_Iter_Sol} and Gaussian elimination are used as the coarse grid solver.

To construct the spatial grid hierarchy $\D_0, \D_1, \dots, \D_L$, we start with a coarse grid $\D_0$ consisting of 512 grid elements (quadrilaterals) and $n_0 = 1122$ degrees of freedom. ($n_\ell$ denotes the number of degrees of freedom in $\D_\ell$.) Grid $\D_\ell$ of freedom is regularly refined twice to obtain grid $\D_{\ell+1}$. We treat this as one spatial refinement step. After every spatial grid refinement step, the number of grid elements is multiplied by 16, i.e., $n_\ell \approx n_0\cdot 16^{\ell}$, see Table~\ref{tab:adaptiveTS_times}). We also construct the temporal grid hierarchy $\Tau_0, \Tau_1, \dots, \Tau_L$. The time step on each temporal grid is denoted by $\tau_\ell$ with $\tau_{\ell+1} = \tfrac{1}{4} \tau_\ell$. The number of time steps on the $\ell$th grid (level) is $r_{\ell+1} = 4 r_\ell$, so that $r_\ell = r_0 4^\ell$, where $r_0$ is the number of time steps in $\Tau_0$. On the $\ell$th level, the MLMC uses the grid $\D_{\ell}\times \Tau_{\ell}$. According to \eqref{eq:1st_ord_conv}, for the numerical solution $\sol_\ell = \sol_{h,\tau}$ on $\D_{\ell}\times \Tau_{\ell}$, we have:
\begin {equation} \label {eq:disc_error}
 \| \conc - \conc_\ell \|_2 = \mathcal{O} (4^{-\ell}).
\end {equation}
Up to four grid levels $\ell$ were used in the numerical experiments.

In the context of this work, it is critical to estimate the numerical complexity of the deterministic solver with respect to the grid level $\ell$. The most time consuming part of the simulation is the solution of the discretised nonlinear system. It is usually difficult to predict the number of Newton iterations in each time step. {\color{black} But in our numerical experiments, two Newton iterations were sufficient to achieve the required accuracy independently on the grid level.} Accordingly, the linear solver was called at most twice per time step. Furthermore, the convergence rate of the geometric multigrid method does not depend on the mesh size (cf.\ \cite{Hackbusch_Iter_Sol}). Thus the computational complexity of a time step is $\mathcal{O} (n_{\ell})$, where $n_{\ell}$ is the number of degrees of freedom at grid level $\ell$. Therefore, the total numerical cost of computing a scenario at grid level $\ell$ for $r_{\ell}$ time steps is
\begin{equation}
\label{eq:CompComplexity}
s_\ell = \mathcal{O} (n_\ell r_\ell), \quad s_\ell \propto s_{\ell-1} \cdot 4^{2}\cdot 4^1= 4^3 \cdot s_{\ell-1}.
\end{equation}
\section{Multilevel Monte Carlo}
\label{sec:MLMC}
To reduce the total computing cost, we apply the MLMC method, which is a natural idea because the deterministic solver uses a multigrid method (see Section~\ref{sec:Num}). The MLMC method efficiently combines samples from various levels. A more in-depth description of these techniques is found in~\cite{MLMC_PDE_anal11,CMLMC,giles2008,giles2015,ErikOptGeom15,teckentrup2013further,Litv_Scattered19}.

%Before to apply the MLMC, we implement and check some preliminary conditions needed for successful MLMC performance.

We let $\xib(\omega)$ and $g(\xib)=g(\xib(\omega))$ represent a vector of random variables and the QoI, respectively, where $\omega$ is a random event. In this work, $g$ is an integral over some of $\Delta_i$'s or $\D$ with the integrand depending on $\sol$, s.\ Section \ref{sec:numerics}. The MLMC method aims to approximate the expected value $\EXP{g}$ with an optimal computational cost. It constructs a telescoping sum, defined over a sequence of spatial and temporal meshes, $\ell=0, \ldots, L$, as described next, to achieve this goal. The QoI $g$, numerically evaluated on level $\ell$, is denoted by $g_{h_{\ell},\tau_{\ell},\ell}$ or, for simplicity, by just $g_\ell$, where $h_{\ell}$ and $\tau_{\ell}$ are the discretization steps in space and time on level $\ell$. Furthermore, we assume that $ \EXP{g_{h,\tau}} \rightarrow \EXP{g}$ as $h\rightarrow 0$ and $\tau \rightarrow 0$.  
%\textcolor{black}{From now on, when we write $\EXP{g}$ or $\EXP{g_{\ell}}$ we mean $\EXP{g_{h,\tau}}$ or $\EXP{g_{h_{\ell},\tau_{\ell},\ell}}$ respectively.}

\begin{remark} Since the problem described in Subsection \ref{subsec:Henry} is non-stationary, the quantities related to its solution $\sol$ below depend on time, too. Examples of time-dependent quantities are $g$, $g_{\ell}$, $\EXP{g_{\ell}}$, $\Var{g_{\ell}}$. To keep our notation simple, we do not explicitly specify the time argument for them. We shall only indicate it where it is essential.
\end{remark}

Let $s_0$ be the computing cost to evaluate one realization of $g_0$ (the most expensive one from all computed realizations on 0th mesh). Similarly, $s_\ell$ denotes the computing cost of evaluating $g_\ell - g_{\ell-1}$. For simplicity, we assume that $s_\ell$ for $g_{\ell} - g_{\ell-1}$ is almost the same as $s_\ell$ for $g_{\ell}$. The number of iterations is variable; thus, the cost of computing a sample of $g_\ell - g_{\ell-1}$ may fluctuate for various realizations.

The MLMC method calculates $\EXP{g_L}\approx \EXP{g}$ using the following telescopic sum: 
\begin{align}
  \EXP{g_L} &= \EXP{g_{0}} + \sum_{\ell=1}^L \EXP{g_{\ell}-g_{\ell-1}} \label{eq:EgL} \\
  &\approx  %= \sum_{\ell=0}^L \EXP{G_{\ell}}
  m_0^{-1}\sum_{i=1}^{m_0} g_{0}^{(0,i)}  + \sum_{\ell=1}^L \left( m_\ell^{-1}\sum_{i=1}^{m_\ell} (g_{\ell}^{(\ell,i)} - g_{\ell-1}^{(\ell,i)} )\right). \ \label{eq:A}
\end{align}
In the above equation, level $\ell$ in the superscript $(\ell, i)$ indicates that independent samples are used at each correction level.
%where $ G_\ell :=g_\ell- g_{\ell-1}$, $ G_0=g_0$.
% \begin{equation}
% \label{eq:EgLapprox}
%   \EXP{g_L} = \EXP{g_{0}} + \sum_{\ell=1}^L \EXP{g_{\ell}-g_{\ell-1}}, %= \sum_{\ell=0}^L \EXP{G_{\ell}}
% \end{equation}
% Note that both $g_\ell$ and $g_{\ell-1}$ are computed using the same input random parameter $\xib$.
% In the telescoping sum~\eqref{eq:EgL}, the expected values in practice are replaced by sample averages, i.e., \begin{equation}
% \EXP{G_\ell} \approx \est G_\ell  =  M_\ell^{-1}\sum_{m=1}^{M_\ell} G_{\ell,m},
% \end{equation}
% where random variable $G_{\ell,m}$ have the same distribution as  %\thicksim 
% $G_{\ell}$ and are independent identically distributed (i.i.d.) samples.
As $\ell$ increases, the variance of $g_{\ell} - g_{\ell-1}$ decreases. Thus, the total computational cost can be reduced by taking fewer samples on finer meshes.

In our numerical experiments $n_{\ell}=16n_{\ell-1}=\ldots=16^{\ell}n_{0}=4^{d\ell}n_{0}$, $d=2$, and $r_{\ell}=4r_{\ell-1}=\ldots=4^{\ell}r_{0}$. 
In the case of uniform, equidistant mesh, we could also write similar formulas for step sizes: $h_{\ell}=h_{\ell-1}\cdot 4^{-1}=h_{\ell-2}\cdot 4^{-2}=\ldots= h_0\cdot 4^{-\ell}$ and $\tau_{\ell}=\tau_0\cdot 4^{-\ell}$. 

We assume that the average cost of generating one sample of $g_{\ell}$ (the cost of one deterministic simulation for one random realization) is
\begin{equation}
s_\ell = \mathcal{O}(n_{\ell}r_{\ell})
= \mathcal{O}(16^{\ell}n_0\cdot 4^{\ell}r_0)
= \mathcal{O}(4^{2\ell}n_0\cdot 4^{\ell}r_0)
= \mathcal{O}(4^{(2+1)\ell \gamma}n_0 r_0)
= \mathcal{O}(4^{\hat{d}\ell \gamma}n_0 r_0),
%=\mathcal{O}(h_{\ell}^{-1} \tau_{\ell}^{-1})=
%=\mathcal{O}\left( \frac{1}{h_0 \tau_0} 4^{2\ell}4^{\ell} %\right)=
%\mathcal{O}\left( \frac{1}{h_0 \tau_0} 4^{3\ell} \right),
\label{eq:workpl}
\end{equation}
where $\hat{d}=d+1=3$ and $\gamma=1$.
%$\tilde{d}=2+1$, $d=2$ is the spatial dimension and $1$ the temporal dimension.
%, and $\gamma=1$ is determined by the computational %complexity of the deterministic solver (ug4).

%For the solver used here we have $d=2$, $\gamma\approx 1$, and $\beta=2$.

%We let $V_{\ell}$ be the variance of $g_{\ell}-g_{\ell-1}$. 

\begin{defn}
Let $Y_{\ell}:=m_{\ell}^{-1}\sum_{i=1}^{m_{\ell}} (g_{\ell}^{(\ell,i)} - g_{\ell-1}^{(\ell,i)})$, where $g_{-1}\equiv 0$, so that
\begin{align}
\label{eq:Yell} 
\EXP{Y_{\ell}}:=
      \begin{cases}
        \EXP{g_0}, & \ell=0 \\
        \EXP{g_{\ell} - g_{\ell-1}}, & \ell>0 \\
      \end{cases}.
\end{align}
Denote by $Y:=\sum_{\ell=0}^L Y_{\ell}$ the multilevel estimator of $\EXP{g}$ based on $L+1$ levels and $m_{\ell}$ independent samples on level $\ell$, where $\ell=0,\ldots,L$.

Furthermore, we denote $V_0 = \Var{g_0}$ and for $\ell \ge 1$, let $V_{\ell}$ be the variance of $g_{\ell} - g_{\ell-1}$: $V_{\ell}:= \Var{g_{\ell} - g_{\ell-1}}$.
\end{defn}

The standard theory states the following facts for the mean and for the variance:
\begin{equation} \label{eq:goal_variance}
 \EXP{Y}=\EXP{g_L}, \qquad \Var{Y}={\textstyle \sum_{\ell=0}^L m_{\ell}^{-1} V_{\ell}}.
\end{equation}
The cost of the multilevel estimator $Y$ is
\begin{equation}
 S := {\textstyle \sum_{\ell=0}^{L} m_{\ell}s_{\ell}}.
\end{equation}

In the following, we repeat the well-known \cite{giles2015} results on the computation of the sequence $m_0, \dots, m_L$. For a fixed variance $\Var{Y} =: \varepsilon^2 / 2$, the cost $S$ is minimized by choosing as $m_{\ell}$ the solution of the optimization problem
{\color{black}
\begin{equation}
\label{eq:goal_function}
\min_{m_0,\ldots,m_{L}}F(m_0,\ldots,m_{L}),
\end{equation}
where $F(m_0,\ldots,m_{L}):=\sum_{\ell=0}^{L} \left( m_{\ell}s_{\ell}+\mu^2 \frac{V_{\ell}}{m_{\ell}}\right)$,} $\mu^2$ is a Lagrange multiplier.
Thus, at the desired set of $m_{\ell}$, the derivatives of $F$ w.r.t. $m_{\ell}$ are equal to zero:
\begin{equation} \label {eq:zero_deriv}
\frac{\partial F(m_0,\ldots,m_{L})}{\partial m_{\ell}}:= s_{\ell}-\mu^2 \frac{V_{\ell}}{m_{\ell}^2}=0.
\end{equation}
Solving the system \eqref{eq:zero_deriv}, we obtain
\begin{equation} \label{eq_m_l_of_mu}
 m_{\ell}^2=\mu^2 \frac{V_{\ell}}{s_{\ell}}, \quad \text{i.e.} \quad
 m_{\ell}=\mu \sqrt{\frac{V_{\ell}}{s_{\ell}}}.
\end{equation}
Taking into account that the variation $\Var{Y}$ is fixed and substituting \eqref{eq_m_l_of_mu} into \eqref{eq:goal_variance}, i.e. $\sum_{\ell=0}^{L}V_{\ell}/m_{\ell} = \varepsilon^2 / 2$, we obtain an equation for $\mu$:
$$
\sum_{\ell=0}^{L}\frac{V_{\ell}}{\mu \sqrt{\frac{V_{\ell}}{s_{\ell}}}} = \tfrac{1}{2} \varepsilon^2.
$$
From this equation, we get $\mu = 2 \varepsilon^{-2} \sum_{\ell=0}^{L} \sqrt{V_{\ell}s_{\ell}}$, and therefore
\begin{equation}
\label{eq:M_ell}
 m_{\ell} = 2 \varepsilon^{-2} \cdot \sqrt{\frac{V_{\ell}}{s_{\ell}}} \cdot \sum_{i=0}^{L} \sqrt{V_{i}s_{i}}.
\end{equation}
For this set of $m_{\ell}$, the total computational cost of $Y$ is
\begin{equation}
\label{eq:total_cost_MLMC}
S = 2 \varepsilon^{-2} \left( \sum_{\ell=0}^L \sqrt{V_{\ell} s_{\ell}}\right)^2.
\end{equation}
For further analysis of this sum, see \cite{giles2015}, p.4.

%The Theorem~1 in \cite{giles2015} lists the conditions when MLMC is faster than the standard MC. In the worst case, when the dominant computational cost is on the coarsest level, the MLMC has the same computational cost as the standard MC.

%!!!
The mean squared error (MSE) is used to measure the quality of the multilevel estimator:
\begin{equation}
\label{eq:MSE}
\MSE:=\EXP{(Y-\EXP{g})^2}=\Var{Y} + \left( \EXP{Y} - \EXP{g} \right)^2,
\end{equation}
where $Y$ is what we computed via MLMC, and $\EXP{g}$ what actually should be computed.
To achieve
$$
 \MSE \le \varepsilon^2
$$
for some prescribed tolerance $\varepsilon$, we ensure that
\begin{equation} \label{eq:bias_error}
 \left( \EXP{Y} - \EXP{g} \right)^2 = (\EXP{g_L-g})^2  \le \tfrac{1}{2} \varepsilon^2
\end{equation}
and
\begin{equation} \label{eq:var_error}
 \Var{Y} \le \tfrac{1}{2} \varepsilon^2.
\end{equation}
The bias error \eqref{eq:bias_error} corresponds to the discretization error \eqref{eq:disc_error} discussed in Subsection \ref{sec:Num}. Later, in the numerical Section we will see that $\EXP{Y} - \EXP{g} = O(4^{-\alpha L})$ with $\alpha \approx 1$. The bias error can be made smaller than $\varepsilon^2 / 2$ by choosing a sufficiently large $L$. Then, for this $L$, we can compute optimal $m_0, \dots, m_L$ by formula in \eqref{eq:M_ell} to provide \eqref{eq:var_error}.

Combining this idea with a sequence of levels of the simulation grids in which the cost increases exponentially with $\ell$ while the weak error $\EXP{g_L-g}$ and multilevel correction variance $V_{\ell}$ decrease exponentially leads to the following theorem (cf. Theorem 1, p.~6 in \cite{giles2015}):
\begin{theorem}
\label{thm:costMLMC}
Consider a fixed $t=t^*$. Suppose positive constants $\alpha,\beta,\gamma > 0$ exist such that $\alpha \geq \frac{1}{2} \min(\beta, \gamma \hat{d})$, and
\begin{subequations}
\label{eq:q1q2}
\begin{align}
    \vert \EXP{g_\ell-g} \vert & \leq c_1 4^{-\alpha\ell} \label{eq:weak_error_model} \\
    V_{\ell} & \leq c_2 4^{-\beta\ell} \label{eq:strong_error_model} \\
     s_{\ell} &\leq c_3 4^{\hat{d}\gamma \ell}.
\end{align}
\end{subequations}
Then, for any accuracy $\varepsilon < e^{-1}$, a constant $c_4>0$ and a sequence of realizations $\{m_{\ell}\}_{\ell=0}^L$ exist, such that
%the following statements are satisfied [cf.~\eqref{eq:goal}]
%$e(\tilde{G}_{\ell})< \tol$
$\MSE < \varepsilon^2$, where $\MSE$ is defined in \eqref{eq:MSE},
%\MSE%:=\EXP{(Y-\EXP{g})^2}< \varepsilon^2,    
%\end{equation*}
%
and the computational cost is
\begin{align}
\label{eq:mlmc_iso_work} 
%S_\varepsilon\left(\hat{Q}^{ML}_{h,\{m_{\ell}\}}\right)\leq
S=
      \begin{cases}
        c_4 {\varepsilon^{-2}}, & \beta > \hat{d}\gamma \\
        c_4 {\varepsilon^{-2} \left(\log(\varepsilon)\right)^2}, & \beta= \hat{d}\gamma \\
        c_4 {\varepsilon^{-\left(2 +\frac{\hat{d}\gamma-\beta}{\alpha}\right)}},  & \beta < \hat{d}\gamma. \\
      \end{cases}
\end{align}
\end{theorem}

This theorem (see also \cite{hoel2014implementation,hoel2012adaptive,charrier2013,MLMC_PDE_anal11,giles2008}) indicates that, even in the worst-case scenario, the MLMC algorithm has a lower computational cost than that of the traditional (single-level) MC method, which scales as $\mathcal{O}(\varepsilon ^{-2-\hat{d}\gamma/\alpha})$. %\textcolor{black}{DELETE? whereas each case in \eqref{eq:mlmc_iso_work} shows a smaller total work}.
%Furthermore, in the best-case scenario presented above, the computational cost of the MLMC algorithm scales as $\Order{\varepsilon ^{-2}}$.
%, i.e. identical to that of the MC method assuming that the cost per sample is $\Order{1}.$ 
%In other words, for this case, the \CMLMC algorithm can in effect remove the computational cost required by the discretization, namely  $\mathcal{O}(\tol^{-d\gamma/q_1})$.
%111!!!
\begin{remark}
In Theorem~\ref{thm:costMLMC}, the factors $c_1$, $c_2$, $c_3$, $c_4$ as well as the exponents $\alpha$, $\beta$ and $\gamma$ may generally depend on the time point $t^{*}$. This makes $L$ and $m_\ell$ time-dependent, too. In this work, we assume that there exist upper bounds $\hat{c}_i$ for the factors: $c_i \le \hat{{c}_i}$, $i = 1, \dots, 4$, as well as the bounds for the exponents: $\alpha \ge \hat{\alpha}$, $\beta \ge \hat{\beta}$, $\gamma \le \hat{\gamma}$ over the whole time interval $[0, T]$ (later, we will see that this assumption is confirmed by our numerical tests). Then the corresponding $L$ and $m_\ell$ computed from \eqref{eq:bias_error} and \eqref{eq:M_ell} are constant in time, too. For simplicity, in what follows, we omit the hats and refer to these bounds.
\end{remark}
\begin{remark}
In \eqref{eq:bias_error} and \eqref{eq:var_error}, $\varepsilon$ is supposed to have the units of QoI $g$ and to have the corresponding meaning. For example, if $g$ is the mass of the salt in a subdomain, measured in kg, then $\varepsilon$ bounds the error of this mass. This raises the question of appropriate scaling of $\varepsilon$.

Let $E_0:=|\EXPt{\EXP{g_0}}|$, where $\EXPt{\cdot}$ is expectation w.r.t. time.
In this work, it makes sense to consider the error relatively to $g$. For this, in \eqref{eq:bias_error} and \eqref{eq:var_error}, we replace $\varepsilon$ by $\varepsilon \cdot E_0$. Equivalently, we can divide (rescale) $\EXP{g_L-g}$ by $E_0$ and $V_i$ by $E_0^2$. Therefore, we get:
\begin{equation}
\label{eq:Egl}
\frac{ | \EXP{g_L-g} |}{E_0} \leq c_1 4^{-\alpha L}.
\end{equation}
Now, to satisfy \eqref{eq:bias_error}, we want $| \EXP{g_L-g} | \leq  \frac{1}{\sqrt{2}} \varepsilon \cdot E_0$. From this inequality, we can estimate $L$:
\begin{equation*}
c_1 4^{-\alpha L} = \frac{\varepsilon \cdot E_0 }{\sqrt{2}}
\end{equation*}
\begin{equation} \label{eq:num_levels}
 L = - \tfrac{1}{\alpha} \log_4 \frac {\varepsilon \cdot E_0} {\sqrt{2} c_1}.
\end{equation}
Equations \eqref{eq:M_ell} and \eqref{eq:total_cost_MLMC} attain the form
\begin{equation} \label{eq:scaled_m_l_S}
 m_{\ell} = \frac {2 \varepsilon^{-2}} {{E_0} ^2} \cdot  \sqrt{\frac{V_{\ell}}{s_{\ell}}} \cdot \sum_{i=0}^{L} \sqrt{V_{i}s_{i}},
 \qquad
 S = \frac {2 \varepsilon^{-2}} {E_0^2} \left( \sum_{\ell=0}^L \sqrt{V_{\ell} s_{\ell}}\right)^2.
\end{equation}
Note that although $E_0$ is an inaccurate approximation of $\EXP{g}$, it is sufficient to be used in \eqref{eq:num_levels}--\eqref{eq:scaled_m_l_S} for scaling purposes. In contrast to $\EXP{g_\ell}$ with $\ell > 0$, $E_0$ can be better estimated by the MLMC method as a large number of samples are computed on the grid level $0$.
\end{remark}

Using preliminary numerical tests (see Fig.~\ref{fig:weak_strong}), we can estimate the convergence rates $\alpha$ for the mean (the so-called weak convergence) and $\beta$ for the variance (the so-called strong convergence), as well as the constants $c_1$ and $c_2$.
% \begin{subequations}
% \label{eq:q1q2}
% \begin{align}
%     \EXP{g_\ell-g_{\ell-1}} & = \mathcal{O}(h_\ell^{q_1}) \label{eq:weak_error_model} \\
%     \var{g_\ell-g_{\ell-1}} &= \mathcal{O}( h_{\ell-1}^{q_2}) \label{eq:strong_error_model} 
% \end{align}
% \end{subequations}
In addition, $\alpha$ is strongly connected to the order of the discretization error (see Section~\ref{sec:Num}), which equals $1$. Note that precise estimates of parameters $\alpha$ and $\beta$ are crucial to distribute the computational effort optimally.
\section{Numerical Experiments}
\label{sec:numerics}
%The goal is to reduce the total computational cost of stochastic simulations. 
%{\color{black} Alexander, say that we consider all the point-boxes, not only no. 9. The no. 9 is presented as the ``most representative case''. The other QoIs are important, too, but misplaced. You'd better place these formulas at the end together with the corresponding graphs.}
%We use the MLMC method to compute the mean value of $\sol(t,\bx)$ in the whole domain.
%We also fix $\bx=\bx_9$ and plot the
%solution $\sol(t,\bx_9)$ at the point $\bx_9:=(1.65, -0.75)$.

In this section, we perform numerical tests with the MLMC method described above. One of our aims is to compare the theoretical predictions with the obtained numerical results. We present two types of numerical experiments. The first type (denoted by A1 - A3) demonstrate the solution in the whole domain, in a point, and salt and fresh water integral values:
\begin{equation}
\label{eq:integral_s}
Q_S(t,\omega):=\int_{\bx\in \D} \sol(t,\bx,\omega) \rho(t,\bx,\omega)  d\bx,
\end{equation}
\begin{equation}
\label{eq:integral_fw}
Q_{FW}(t,\omega):=\int_{\bx\in \D} I(\sol(t,\bx,\omega) \le 0.012178) d\bx,
\end{equation}
where $I(\cdot)$ is the indicator function identifying a subdomain $\{\bx:\; \sol(t,\bx,\omega) \le 0.012178\}$.
For these integrals, we observe numerically $\alpha \approx 2$ (see \eqref{eq:q1q2}) that agrees with the theoretical value. The idea of these tests is to collect more information about the problem.

{\color{black}
\begin{remark}
Our motivation for the threshold $0.012178$ (in \eqref{eq:integral_fw}) is technical: it was used in our simulations of coastal aquifers at the North Sea \cite{SWLRHEGW-SaltwaterInNorthSea2018}, where $1m^3$ of saltwater contains 35kg of salt (corresponding to $\sol=1$ in the model) and has a density of $1035 \frac{kg}{m^3}$. Then the maximum recommended salinity of drinking water $412\frac{mg}{l}$ corresponds to the scaled mass fraction $\sol = 0.012178$. The choice of this value is not essential.

\end{remark}
}

In the second type (denoted by B1-B4), as a QoI, we consider an integral over a small sub-regions around a point (see the list of points in \eqref{eq:12points}):
\begin{equation}
\label{eq:integral_box}
Q_i(t,\omega):=\int_{\bx\in \Delta_i} \sol(t,\bx,\omega) \rho(t,\bx,\omega)  d\bx, \quad \Delta_i = [x_i-0.1,x_i+0.1]\times [y_i-0.1,y_i+0.1],\; i=1,\ldots,15.
\end{equation}
The value of $Q_i$ is the mass of the salt in a subdomain $\Delta_i$. The size of each $\Delta_i$ is small ($0.2^2=0.04$), compared to $\D$.
%NO!According to our observations, all $Q_i$ demonstrate similar behaviour w.r.t. $t$ and $\omega$. 
We choose the point with index $9$ (see Fig.~\ref{fig:Henry2d-scheme_points}) and the corresponding domain $\Delta_9$ as an example.
The numerical scheme provides only the first order of the accuracy to compute $Q_i(t,\omega)$, i.e., the convergence rate $\alpha$ (weak convergence) should be $\approx 1$ in \eqref{eq:weak_error_model}.

%meaning the mass of the fresh water at a time $t$. %The computed QoIs are compared with those computed using the QMC approach. 
%Each simulation may contain up to $n=0.5\cdot 10^6$ spatial mesh points and a few thousand time steps ($r=6016$ on the finest mesh).

%
% In the following, we describe a numerical example, where we assume that the porosity coefficient and the recharge are uncertain.
% The mean and the variance of $\sol(t,\bx,\thetab)$ is computed by the MLMC method. 
% The reference solution is estimated by the QMC method (Halton sequence).

\textbf{Uncertain porosity and recharge:} 
{\color{black}
Often unknown porosity is modelled by a random field. This random field is approximated by a truncated Karhunen-Lo\'eve expansion (KLE). KLE requires knowledge of the covariance matrix, which is typically assumed to be from the very large Mat\'ern class of functions. The KLE components are computed by solving an auxiliary eigenvalue problem and assuming a random distribution for the random variables. This approach is difficult and requires a separate article to describe it. We have implemented a simple alternative. We used $L_2$ orthogonal functions ($\sin()$ and $\cos()$), which mimic the orthogonal functions used in KLE. The uniform random variables used are also very common, usually one of three options is chosen: uniform, Gaussian or log-normal.}

Additionally, we assume the presence of two horizontal layers: $y\in (-0.8,0]$ (the upper layer) and $y\in [-1, -0.8] $ (the lower layer). The porosity inside each layer is uncertain and is modeled as in \eqref{eq:poro_2levels}:
%
%\begin{align}
%  \poro(\bx,\xib) &= 0.35\cdot(1+0.15(\xi_2\cos(\pi x/2) + \xi_2 \sin(2\pi y)+\xi_1 \cos(2\pi  x)))\cdot C_0(\xi_1), \label{eq:poro_2levels} \\
%  \text{where}\;\;C_0(\xi_1) &= \left\{ 
%  \begin{array}{ll}
%    1 + 0.2  \xi_1 & \text{if}\; y<-0.75\\
%    1 - 0.2  \xi_1 & \text{if}\; y\geq -0.75,
%    \end{array} 
%  \right. \label{eq:poro_2levels_C}
%\end{align}
\begin{align}
  \poro(\bx,\xib) &= 0.35 \cdot C_0(\xi_1) \cdot C_1(x,y,\xi_1,\xi_2) \cdot C_2(x,y,\xi_1,\xi_2), \label{eq:poro_2levels} \\
  \text{where}\;\;C_0(\xi_1) &= \left\{ 
  \begin{array}{ll}
    1.2 \cdot (1 + 0.2  \xi_1) & \text{if}\; y<-0.8\\
    1 & \text{if}\; y\geq -0.8,
    \end{array} 
  \right. \label{eq:poro_2levels_C} \\
  C_1(x,y,\xi_1,\xi_2) &= 1 + 0.15 (\xi_2 \cos(\pi x/2) - \xi_2 \sin(2\pi y) + \xi_1 \cos(2\pi  x)), \\
  C_2(x,y,\xi_1,\xi_2) &= 1 + 0.2 (\xi_1 \sin(64 \pi x) + \xi_2 \sin(32 \pi y)).
\end{align}
Random fields generated in this way provide: a) periodicity of high and low porosity areas in the aquifer; b) multi-scale behaviour; c) spatial dependence. This approach has been agreed with the developers of the UG4 software, who have many years of experience in solving very realistic problems at an industrial level. The recharge flux is also uncertain and is equal to
%\begin {eqnarray}
% \hat{q}_{\mathrm in} = -6.6\cdot 10^{-2}(1+0.5\cdot \xi_3),
%\end {eqnarray}
\begin {eqnarray} \label {e_recharge_of_param}
 \hat{q}_{\mathrm in} (t, \xi_3) = -6.6\cdot 10^{-2} (1 + 0.5 \cdot \xi_3) (1 + \sin \tfrac{\pi t}{40}).
\end {eqnarray}
Here $\xi_1$, $\xi_2$, and $\xi_3$ are sampled independently and uniformly in $[-1,1]$.

Note that there are many ways to model recharge, and we chose to model it by a function of $t$ with one random variable $\xi_3$. Of course, one can use more random variables to model the recharge, but we are limited by the available computational resources (i.e., by the number of independent random variables which we can use). Therefore, we decided to use only one RV. Taking into account a periodic nature of precipitations, we decided to model it by a periodic function, for example, by $\sin(\cdot)$.

{\color{black}
In the settings of the classical Henry problem, the flux $\hat{q}_\mathbf{in}$ is simply prescribed. It models the flux of the water coming from the ``land'' (due to rivers, precipitation etc.) to the coast. This means in the real situation, on the one hand, $\hat{q}_\mathbf{in}$ should depend on the hydrogeological parameters of the medium further on the left, behind the left boundary. And for these parameters, the same stochastic modeling should be applied. Furthermore, this flux should depend on the flow inside the domain, i.e. on the porosity and the permeability in the problem setting, and not only near the left boundary. These arguments reflect the general difficulties with imposing realistic in- and outflow boundary conditions but their discussion is beyond the scope of our paper. This is a modeling issue that must be considered aside of the numerical methods.
}

{\color{black}
Note that the porosity (\ref{eq:poro_2levels}) and therefore the permeability (\ref{e_perm_Kozeny_Carman}) are highly oscillating functions of the geometric coordinates $(x, y)$. Input interfaces of some codes assume precalculation of these values and storing them in e.g. vertex- or cell-centered rasters. During assembling of the discretized algebraic systems, these quantities are interpolated at the integration points of the finite-volume or finite-element schemes. For multilevel approaches, these points depend on the grids in the hierarchy, and the resolution of the rasters as well as the type of the interpolation can influence the properties of the methods. In constrast to that, our implementation does not involve precalculation of the porosity and the permeability fields, so that the values (\ref{eq:poro_2levels}) and (\ref{e_perm_Kozeny_Carman}) are directly evaluated at the integration points during the assembling phase. The same holds for the recharge (\ref{e_recharge_of_param}) as a function of time.
}

\begin{figure}[t!]
\begin {center}
\includegraphics[width=0.32\textwidth]{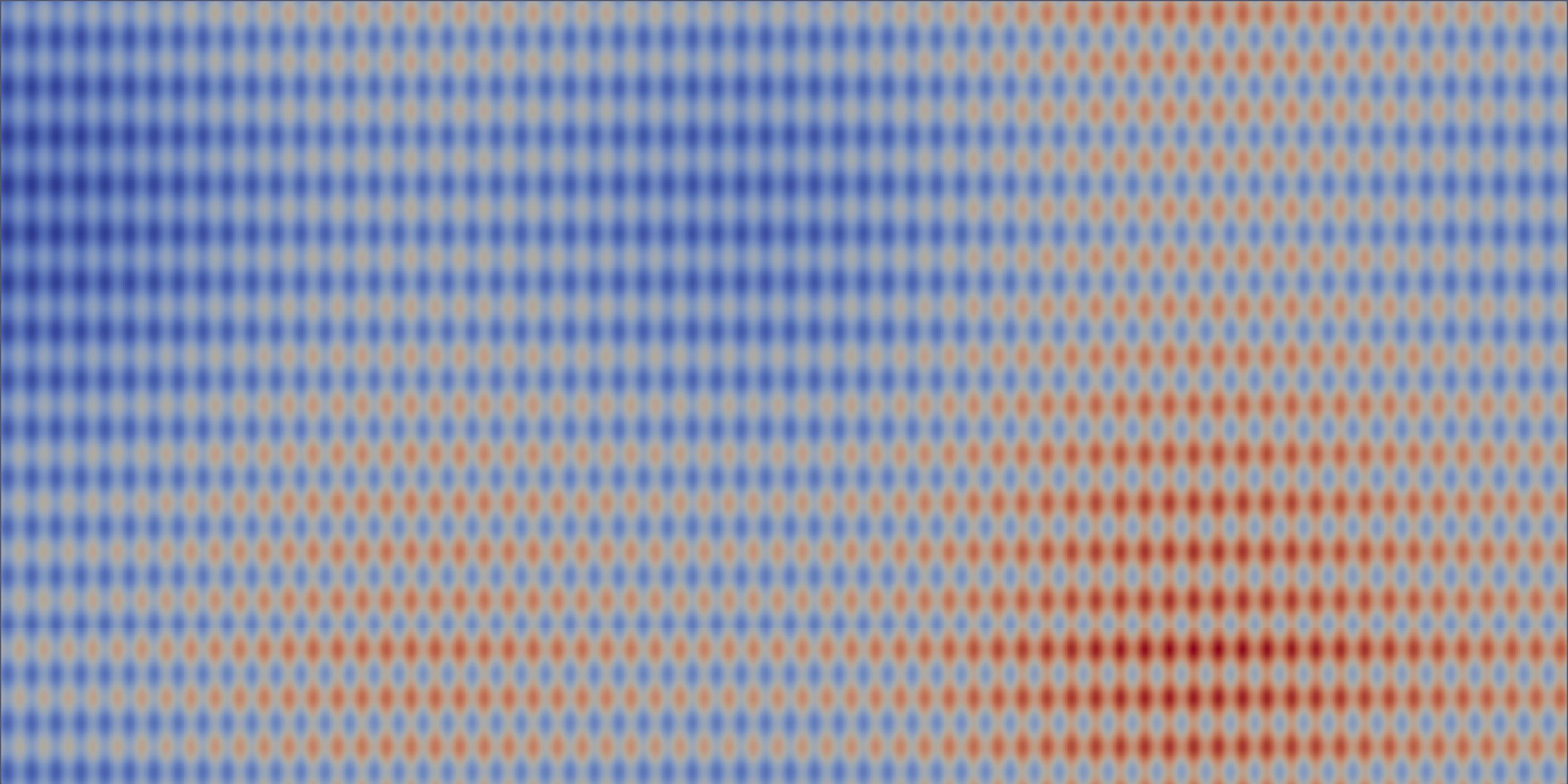}\,
\includegraphics[width=0.32\textwidth]{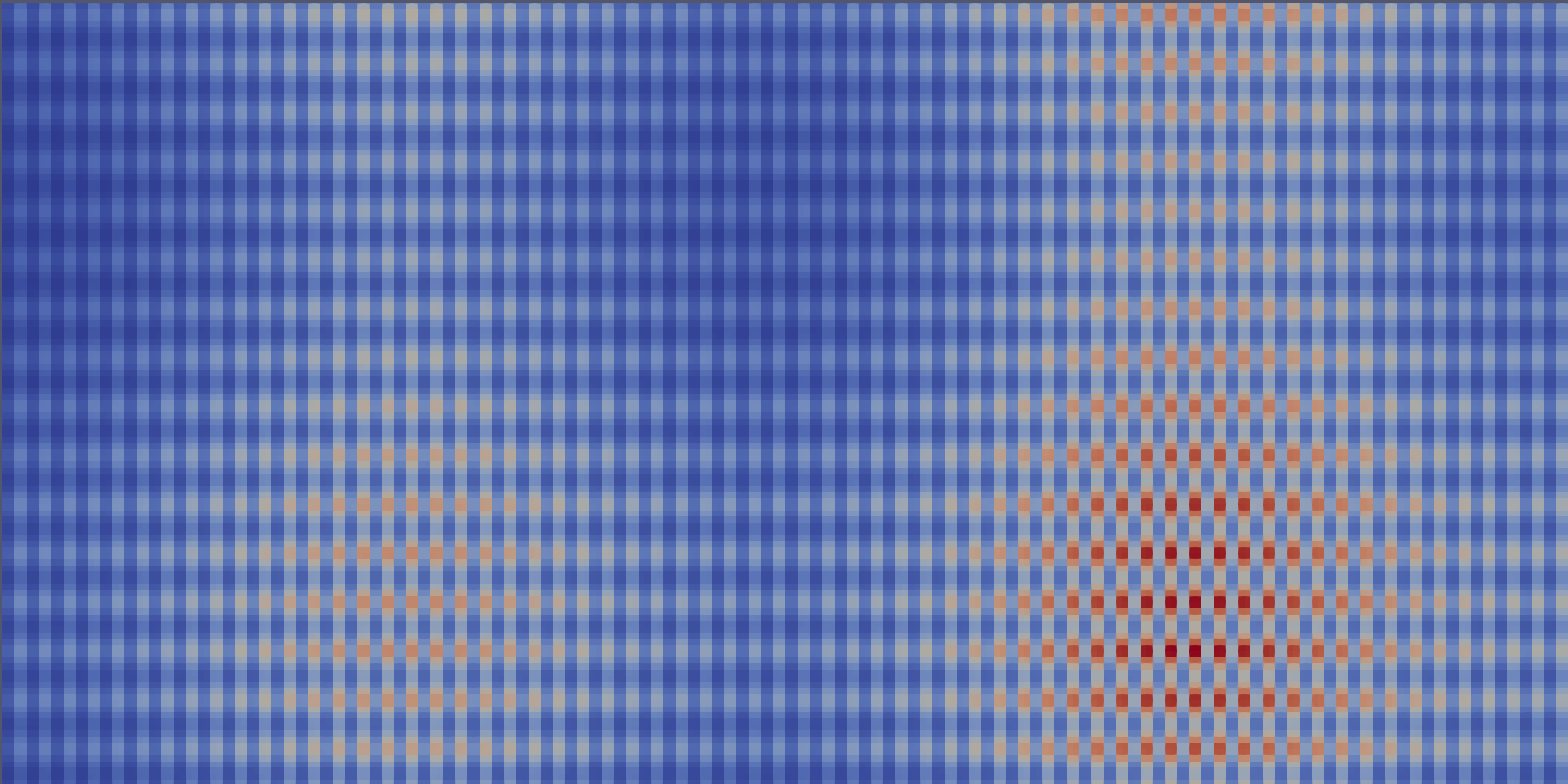}\,
\includegraphics[width=0.32\textwidth]{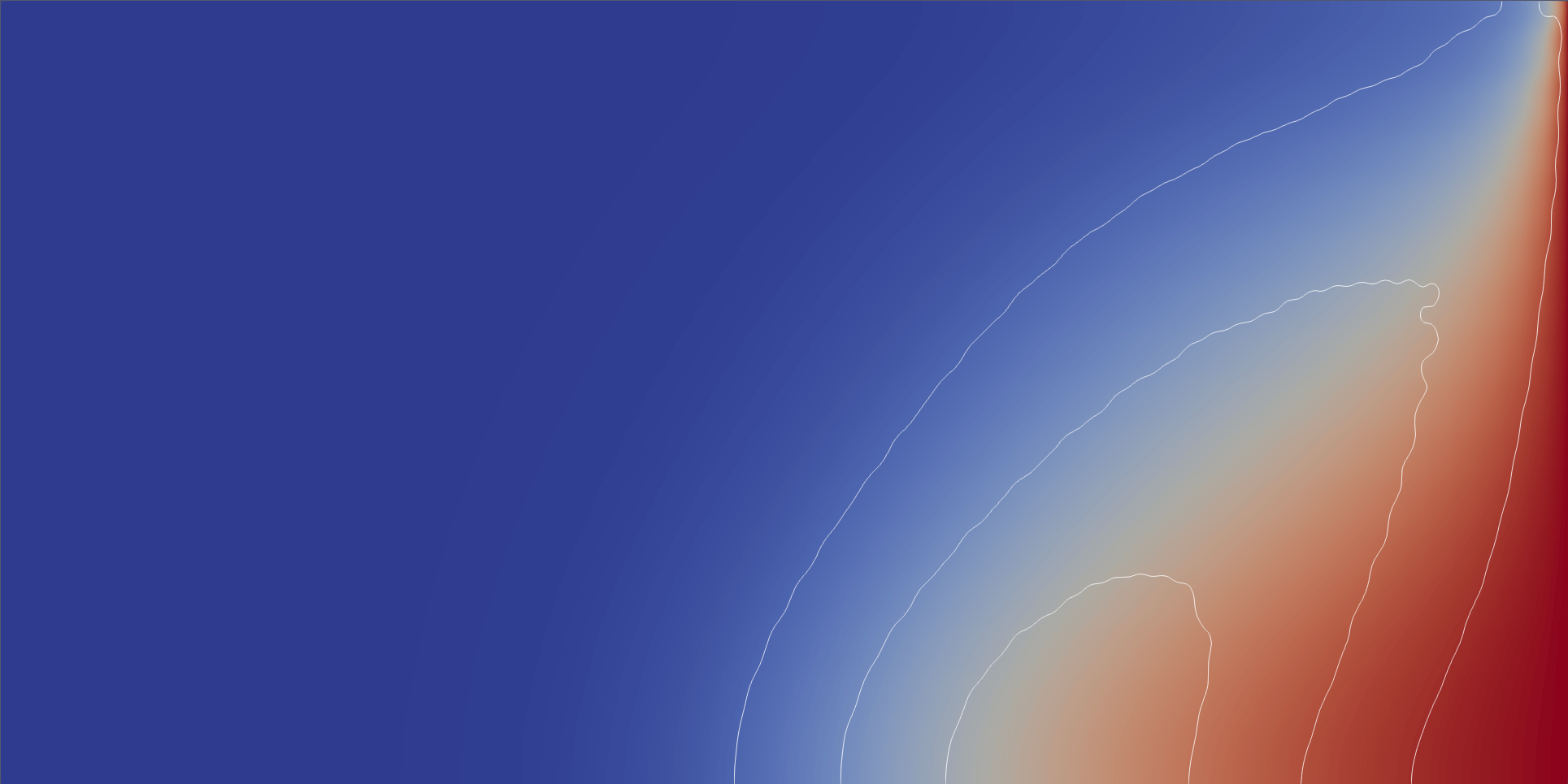}\\
\;
% \includegraphics[width=0.49\textwidth]{figs/fihemoca600-lev7-t47-p300-poro_0_248_0_499.png}%
% \;
% \includegraphics[width=0.49\textwidth]{figs/fihemoca600-lev7-t47-p300-c-adiff_0_1_0_4.png}%
%
\caption {(left and center) A realisation of porosity $\phi(\xib^*) \in [0.18, 0.59]$ and permeability $K \in [1.77e-10, 4.35e-9]$. (right) Corresponding mass fraction $\sol(T,\bx,\phi(\xib^*)) \in [0,0.35]$ with isolines $\{\bx:\; |\sol(T,\poro(\xib^*)) - \overline{\sol}(T)|=0.1\cdot i\}$, $i=1,2,3$, $t=T=6016$ s.} % $\xib^*=(0.5898, 0.7257, 0.9616)$,
\label{fig:sol_poro1}
\end{center}
\end{figure}

Figure~\ref{fig:sol_poro1} depicts a random realization of the porosity random field $\poro(\xib)$ for some $\xib$ (left) and the corresponding solution $\conc(t,\bx,\xib)=\sol(t,\poro(\xib))$ at $t=T$ (right). Additionally, four isolines $\{\bx:\; |\sol(T,\poro(\xib)) - \overline{\sol}(T)|=0.1\cdot i\}$, $i=1,2,3,4$, are presented on the right, $T =6016$ $[s]$. The isolines demonstrate the absolute value of the difference between the computed realization $\sol(t,\poro(\xib))$ and the expected value $\overline{\sol}(t)$.
%These computations were performed for $\xib=\xib^*=(0.5898, 0.7257, 0.9616)$ and $t = T=6016$ s. 

\textbf{Software and parallelization:}
The computations presented in this work were performed using the ug4 simulation software toolbox \cite{ug4_HLRS2012, ug4_ref1_2013, ug4_ref2_2013}. This software has been applied for subsurface flow simulations of real-world aquifers (cf.~\cite{SWLRHEGW-SaltwaterInNorthSea2018}). The toolbox was parallelized using MPI, and the presented results were obtained on the Shaheen II cluster provided by the King Abdullah University of Science and Technology. Every sample was computed on 32 cores of a separate cluster node. Each simulation (scenario) was localized to one node to reduce the communication time between nodes. All scenarios were concurrently computed on different nodes. A similar approach was used in \cite{LitLog3D_20,Litvinenko-UQ-2021}. Simulations times for one realization on different grid levels $\ell$ are presented in Table~ \ref{tab:adaptiveTS_times}.
%were performed on different meshes; thus, the computation time of each simulation varied over a wide range (see Table~ \ref{tab:adaptiveTS_times}). 
%The nodes where the processes have been completed were immediately released and made available for other users.

\subsection{Computation via QMC}
The following calculations are performed using the QMC method with 600 samples (Halton sequence). These numerical tests are not directly relevant to the MLMC method, but are used to better understand the uncertainties in different QoIs.
% Before to do numerical text with differences $g_{\ell} - g_{\ell-1}$, we would like to do some more auxiliary numerical tests with $g_{\ell}$.

\begin{figure}[htbp!]
\begin{center}
  \includegraphics[width=0.49\textwidth]{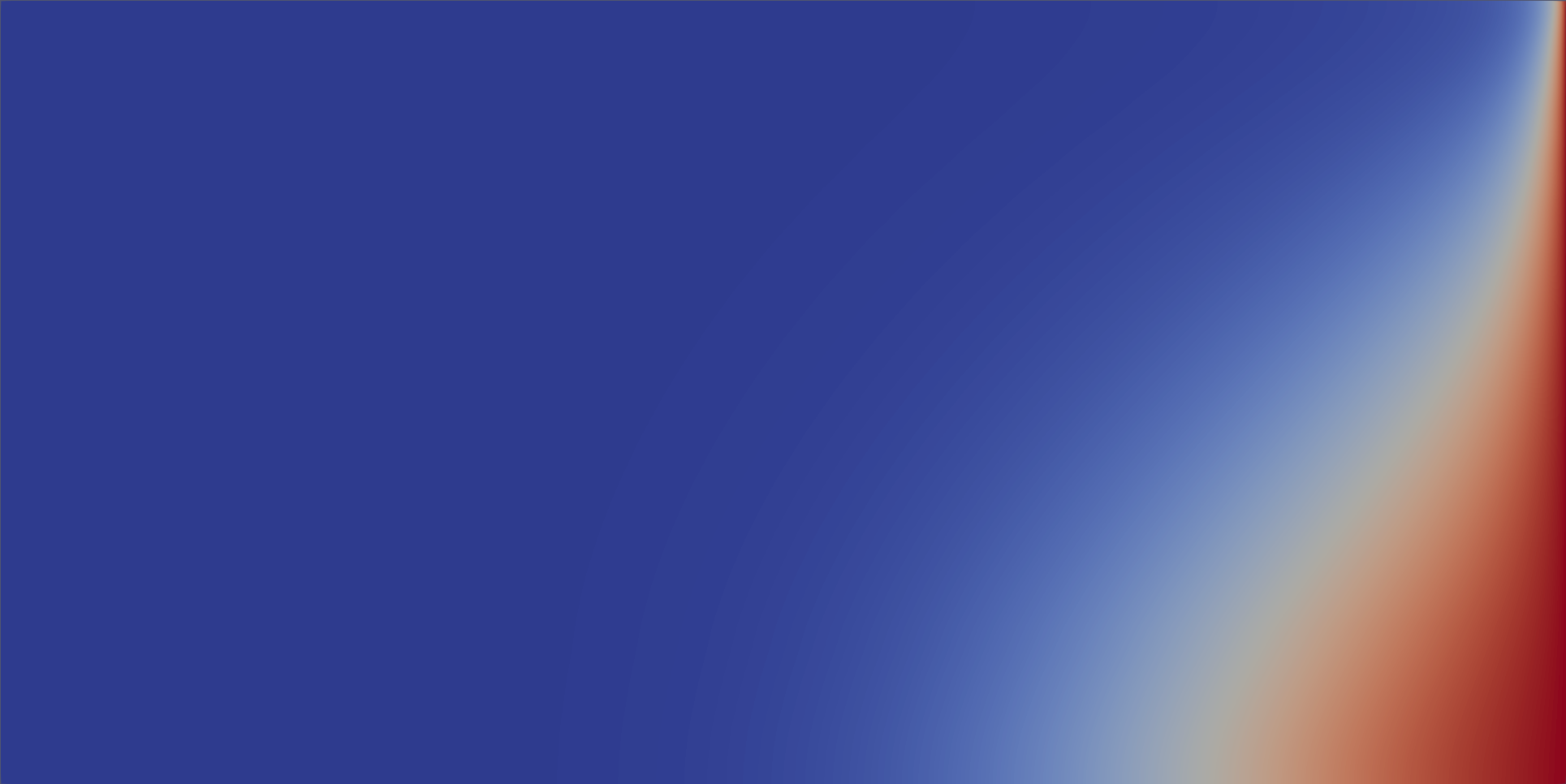}\;
  \includegraphics[width=0.49\textwidth]{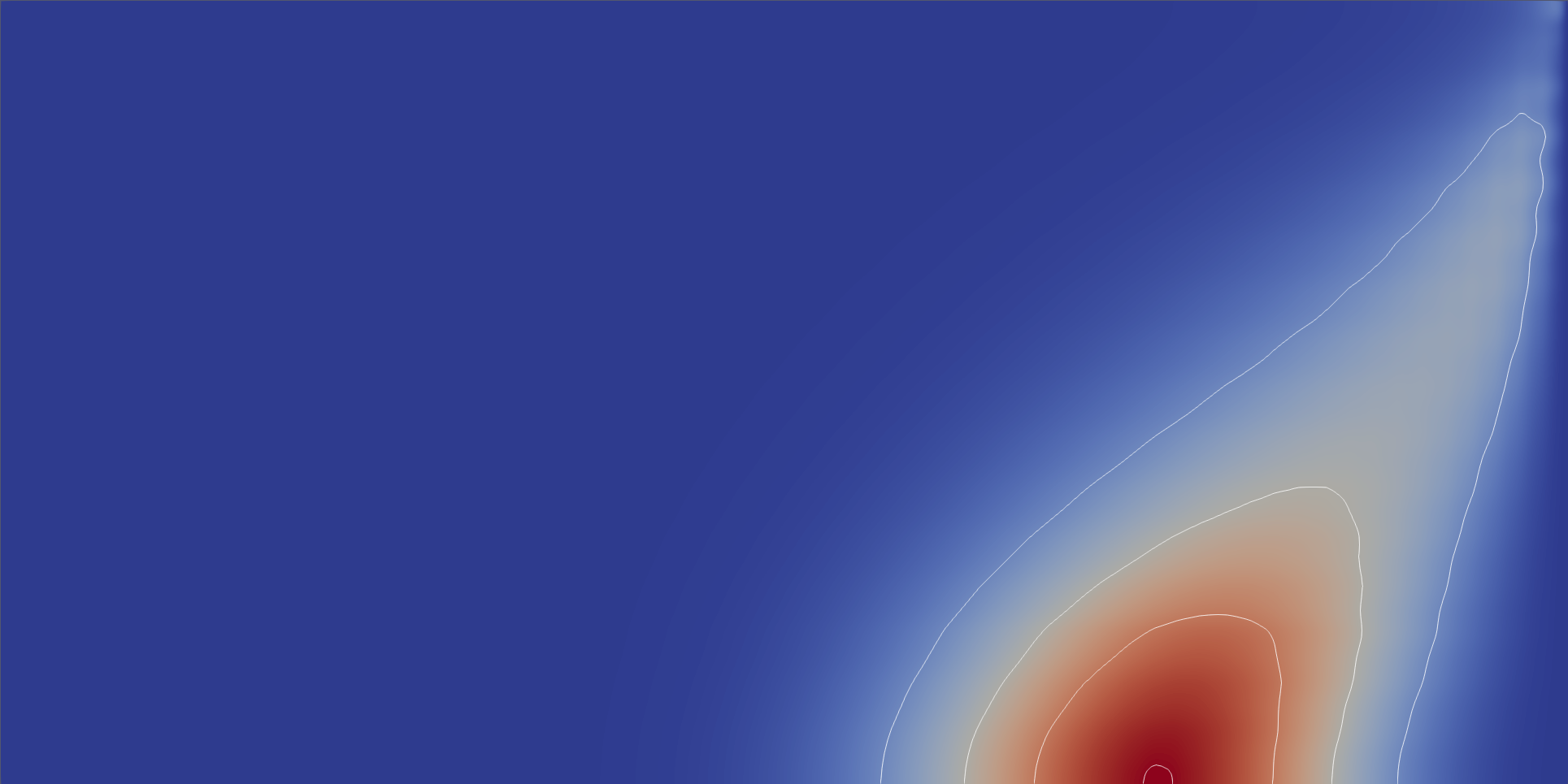}\;
  \caption{(left) Mean value $\overline{\sol} \in [0,1]$ and (right) variance $\var{\sol} \in [0.0,0.04]$ of the mass fraction, with contour lines $\{\bx:\; \var{ \sol }=0.01\cdot i\}$, $i=1..3$, $t=T=6016$ s.}
    \label{fig:countours_mean_var}
\end{center}    
\end{figure}

\textbf{Test A1.} The mean and variance of the mass fraction are provided in Fig.~\ref{fig:countours_mean_var} on the left and right, respectively. The expectation takes values from $[0,1]$. The value $0$ corresponds to the blue colour and  the value $1$ to the dark red. The variance range is $[0,0.04]$. The areas with high variance (dark red) indicate regions with high variability/uncertainty. Such regions may need additional attention from specialists (e.g., placement of sensors). The right image displays four contour lines
  $\{\bx:\; \var{ \sol }(t,\bx)=0.01\cdot i\}$, $i=1..4$, $t=T=6016$.

We observed that the variability (uncertainty) of the mass fraction can vary from one grid point to another. At some points (dark blue regions) the solution does not change. At other points (white-yellow regions) the variability is low (light red regions) or high (dark red regions). In regions of high uncertainty, it is useful to refine the mesh and apply the MLMC method.

\begin{figure}[htbp!]
\begin{center}
  \includegraphics[width=0.49\textwidth]{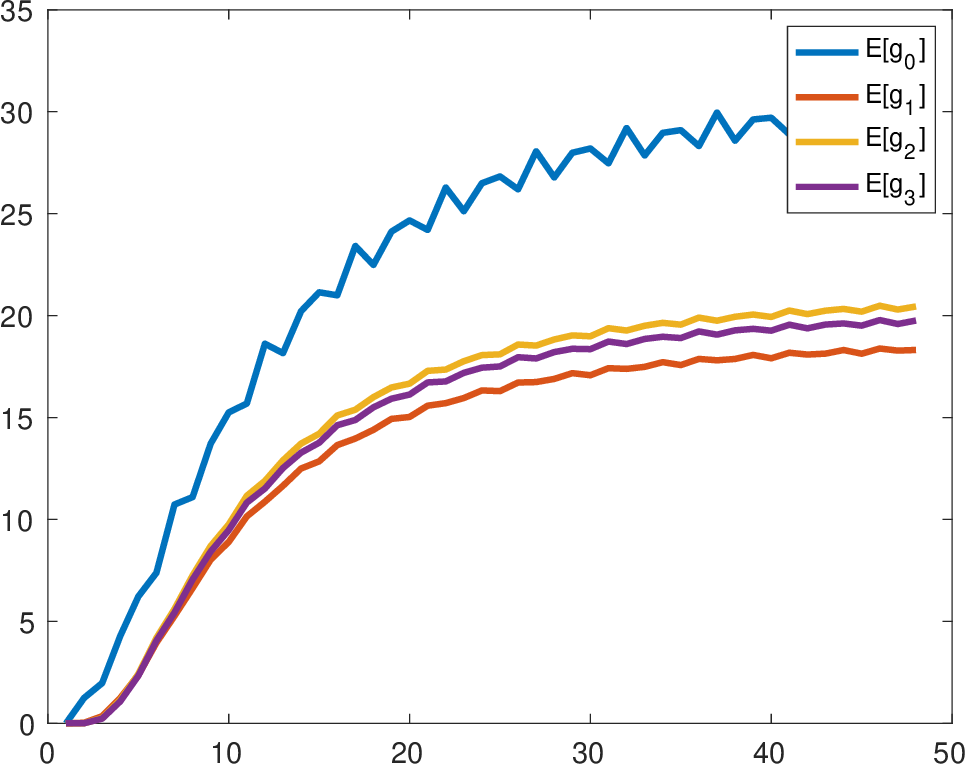}\;
  \includegraphics[width=0.49\textwidth]{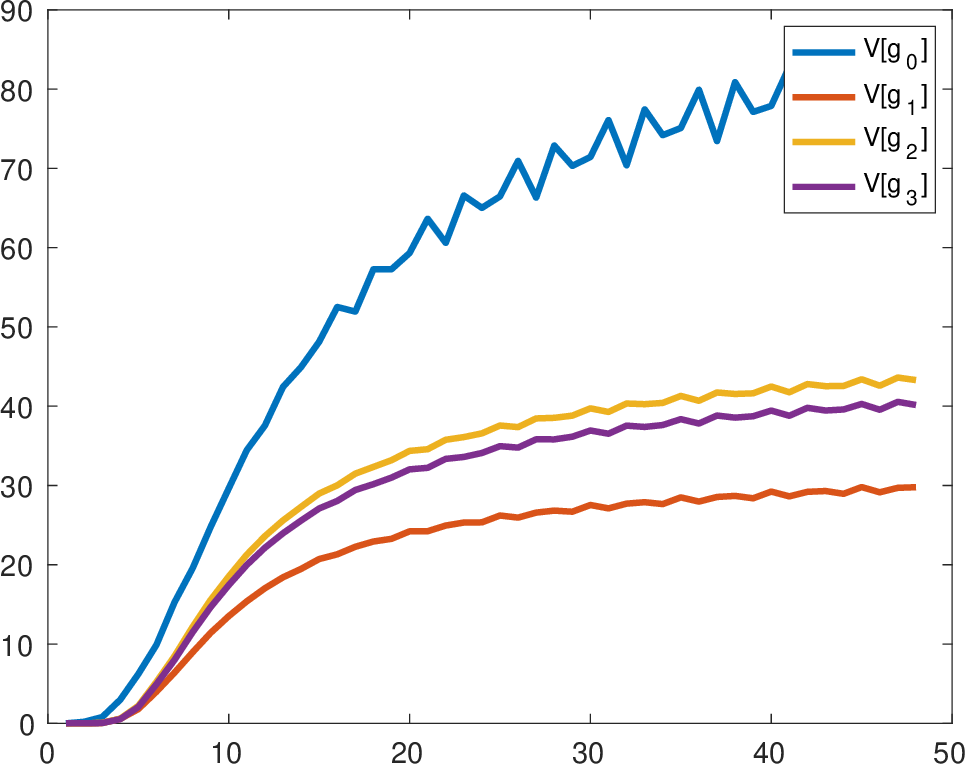}\;
  \caption{(left) Mean values $\EXP{{\sol}(t,x_9,y_9)}$ and (right) variances $\var{\sol}(t,x_9,y_9)$ of the mass fraction computed on levels 0,1,2,3 vs. time $t$.}
    \label{fig:mean_var_levels}
\end{center}    
\end{figure}

\textbf{Test A2.} As a further numerical test, we want to know the range in which the mean and variance of the QoI $g_{\ell}$ change. In Fig.~\ref{fig:mean_var_levels} we visualise the mean $\EXP{{\sol}(t,x_9, y_9)}$ (on the left) and the variance $\var{\sol}(t,x_9, y_9)$ (on the right) at the point $(x_9,y_9)$, computed on mesh levels $\ell=0,\ldots,3$. So, we see that the mean changes in the range $\approx [0, 30]$ and the variance in $\approx [0,80]$.
{\color{black}
Small oscillations (blue line) in the values $\EXP{g_0}$ and $\Var{g_0}$ are due to the oscillations in the recharge, $\sin()$ function in \eqref{e_recharge_of_param}.
}
This experiment gives us an idea of how to choose $\varepsilon^2$ for the MSE error.

%
%\newpage

\begin{figure}[htbp!]
\begin{center}
   \includegraphics[width=0.48\textwidth]{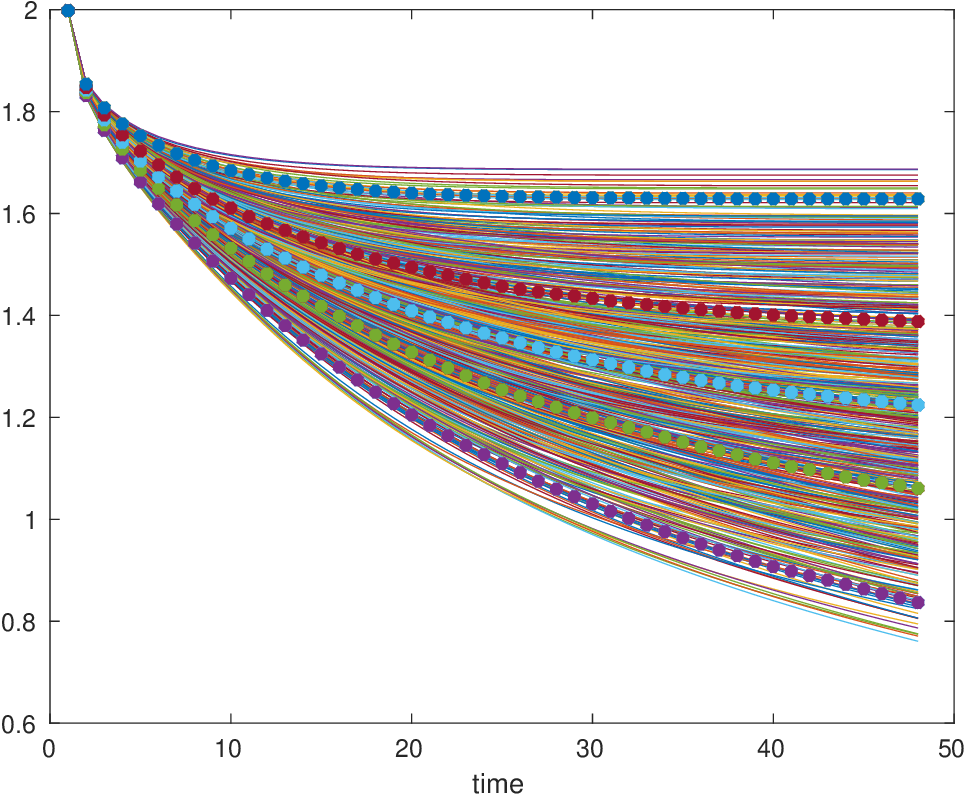}\;
   \includegraphics[width=0.48\textwidth]{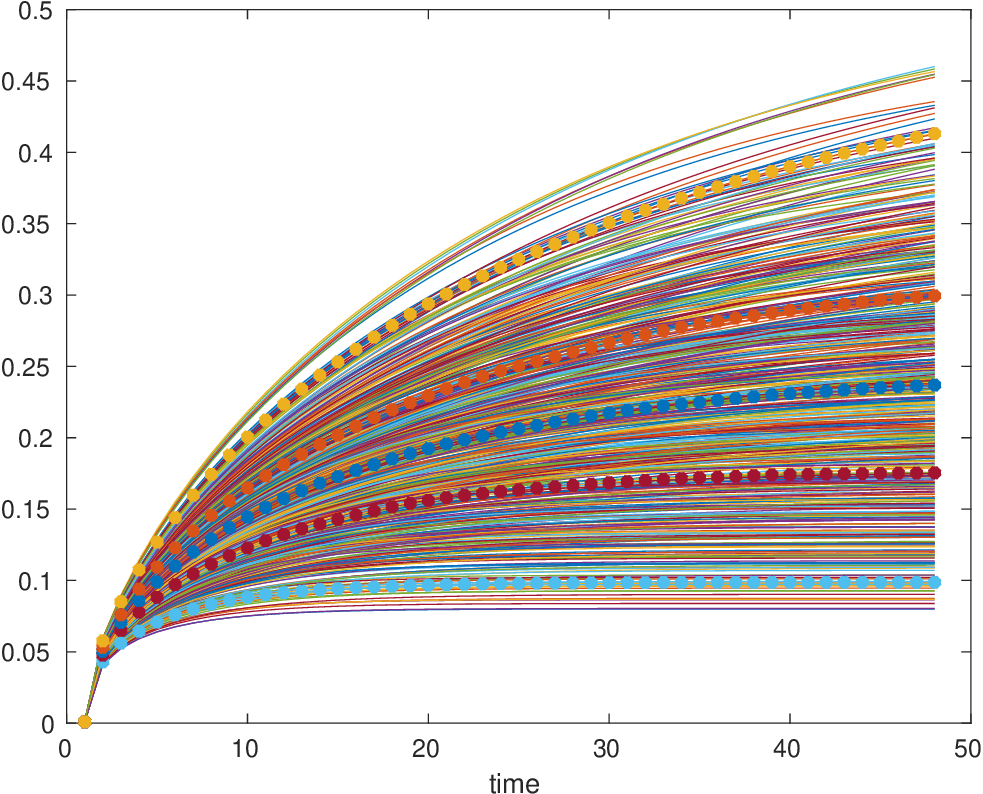}\;
\caption{Six hundred realisations of the integral values $Q_{FW}(t)$ (left) and $Q_{S}(t)$ (right). The $x$-axis represents the time $t=1\tau,\ldots,48\tau$; dotted curves denote five quantiles: 0.025, 0.25, 0.50, 0.75 and 0.975 from bottom to top.}
 \label{fig:600pure_water_realisations}
\end{center}
%Matlabcode/exceed_prob2_pure.m
\end{figure}

\textbf{Test A3.}  600 realisations of $Q_{FW}(t)$ (left) and $Q_{S}(t)$ (right) are shown in Fig.~\ref{fig:600pure_water_realisations}.
Time is along the $x$ axis, $t\in[\tau,48\tau]$. In addition, five quantiles are represented by dotted curves from bottom to top and are 0.025, 0.25, 0.50, 0.75 and 0.975 respectively. These plots can be used to identify worst case scenarios for freshwater reserves.

\subsection{Numerical tests for the MLMC method}
\begin{figure}[htbp!]
\begin{center}
  \includegraphics[width=0.49\textwidth]{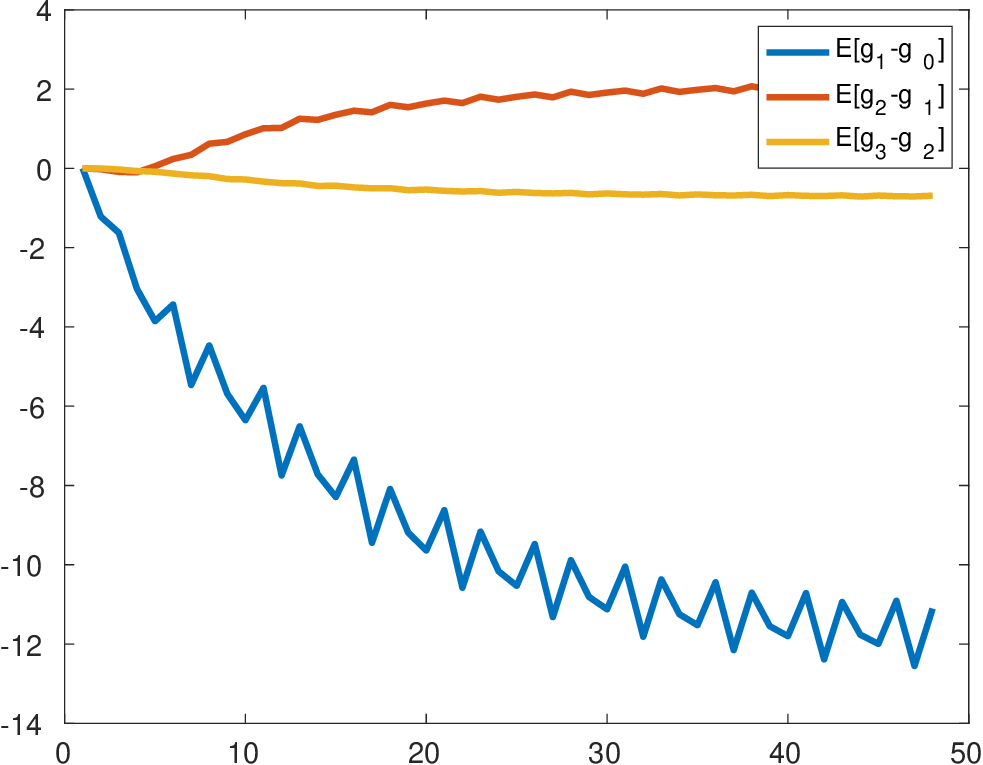}\;
  \includegraphics[width=0.49\textwidth]{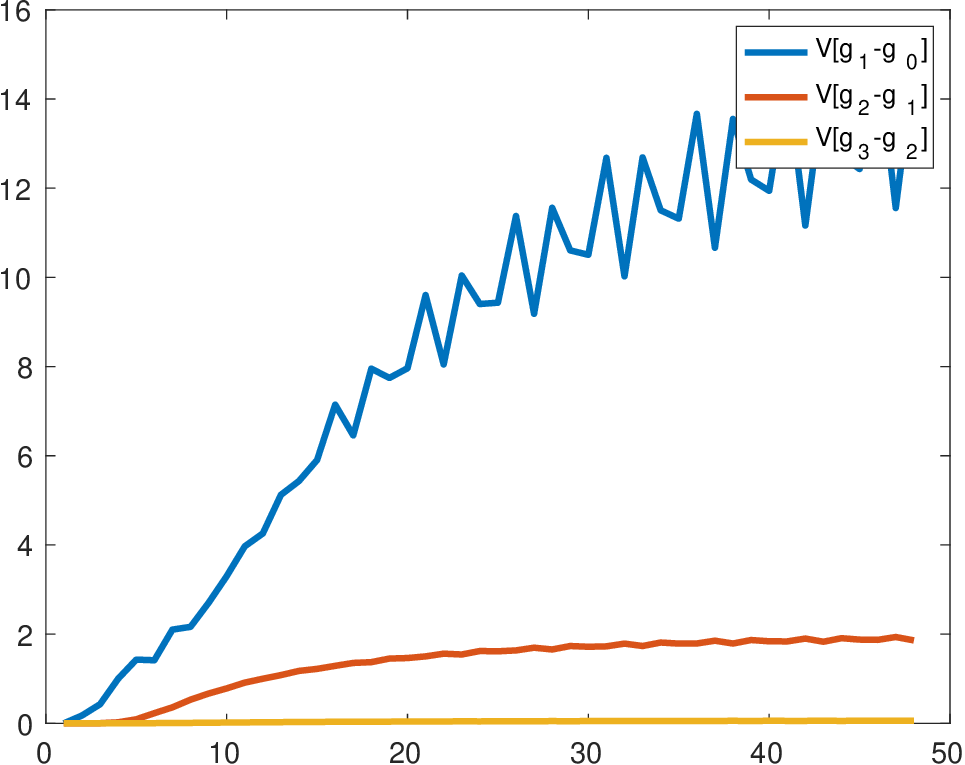}\;
\end{center}
 \caption{(left) The mean value $\EXP{g_{\ell}-g_{\ell-1}}$ and (right) the variance value $\Var{g_{\ell}-g_{\ell-1}}$ as a function of time for $t\in[\tau, 48\tau]$, $\ell=1,2,3$.}
 \label{fig:mean_var_decay_levels}
 %dima_Henry_points_pdfs_diffs_6Apr.m
\end{figure}

The following numerical experiments are required to demonstrate the work and efficiency of the MLMC method.\\
\textbf{Test B1.}
The next QoI is the integral value as in \eqref{eq:integral_box}.
% \begin{equation}
% \label{eq:integral_Qs}
% Q_9(t,\omega):=\int_{\bx\in \Delta_9} \sol(t,\bx,\omega) \rho(t,\bx,\omega)  d\bx.
% \end{equation}
This integral can be estimates from above
\begin{equation}
\label{eq:integral_Qs}
\vert Q_9(t,\omega) \vert \leq \int_{\bx\in \Delta_9} \vert \sol(t,\bx,\omega)\vert\cdot \vert \rho(t,\bx,\omega)\vert  d\bx\leq 0.2^2 \cdot 1000 \cdot 1 = 40,
\end{equation}
where the density $\vert \rho(t,\bx,\omega)\vert \leq 1000$, $\vert \sol(t,\bx,\omega) \vert \leq 1$, and volume of $\Delta_9$ is $0.2^2$.
%The computational subdomain is $\D_s:=\D_9:=[x_9-0.1,x_9+0.1]\times [y_9-0.1, y_9+0.1]$, where $(x_9,y_9)$ is the 9-th measurement point.
Figure~\ref{fig:mean_var_decay_levels} (left) shows the mean value $\EXP{g_{\ell}-g_{\ell-1}}$ as a function of time for $t\in[\tau, 48\tau]$. Figure~\ref{fig:mean_var_decay_levels} (right) shows the variance value $\Var{g_{\ell}-g_{\ell-1}}$ as a function of time for $t\in[\tau, 48\tau]$. In both figures, as we move to finer and finer meshes, we can see that the amplitude decreases, i.e. the accuracy increases.

%\begin{figure}[htbp!]
%\begin{center}
  % \includegraphics[width=0.49\textwidth]{figs/concentr_L4_5_RanRechargeRanPoro.jpg}\;
  % \includegraphics[width=0.49\textwidth]{figs/concentr_L5_6_RanRechargeRanPoro.jpg}\\
  % \includegraphics[width=0.49\textwidth]{figs/concentr_L6_7_RanRechargeRanPoro.jpg}\;
  % \includegraphics[width=0.49\textwidth]{figs/concentr_L7_8_RanRechargeRanPoro.jpg}\\
  % \includegraphics[width=0.49\textwidth]{figs/concentr_L8_9_RanRechargeRanPoro.jpg}
  % \caption{Differences between mass fractions $\sol$ computed at the point $(1.60, −0.95)$ on levels a) 1 and 0, b) 2 and 1 (first row), c) 3 and 2, d) 4 and 3 (second row), and e) 5 and 4 (third row) for 100 realizations ($x$-axis represents time).}
%    \label{fig:diffs_levels}
%\end{center}
%dima_compare_MC_with_MLMC.m
%\end{figure}

\begin{figure}[htbp!]
\begin{center}
  \includegraphics[width=0.32\textwidth]{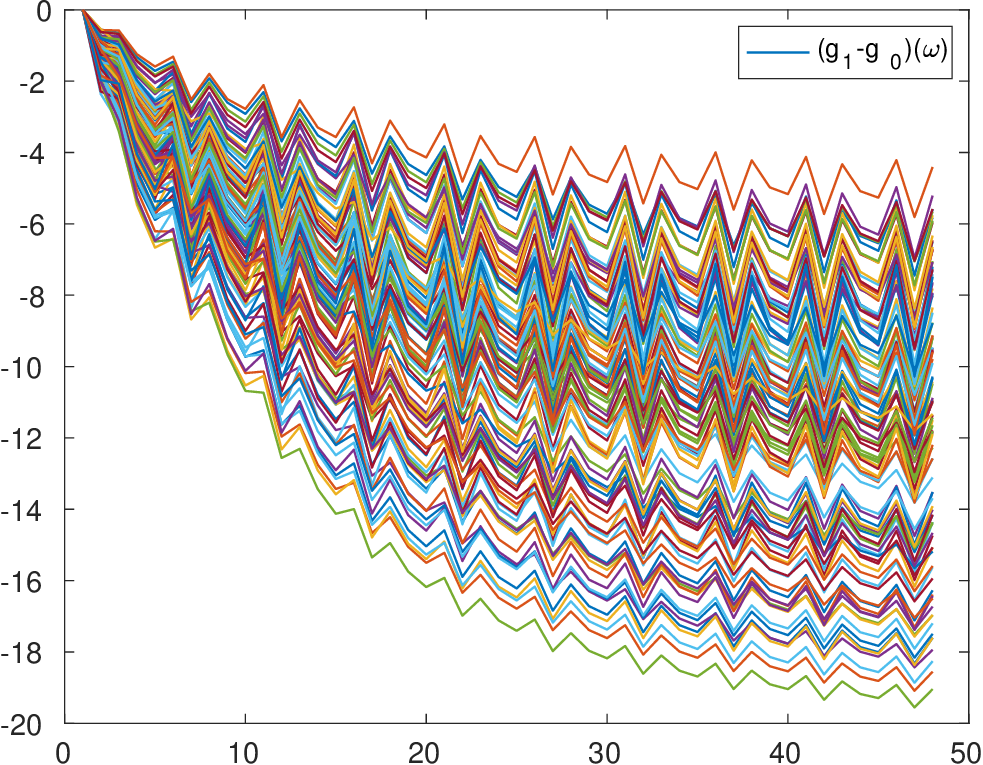}\;
  \includegraphics[width=0.32\textwidth]{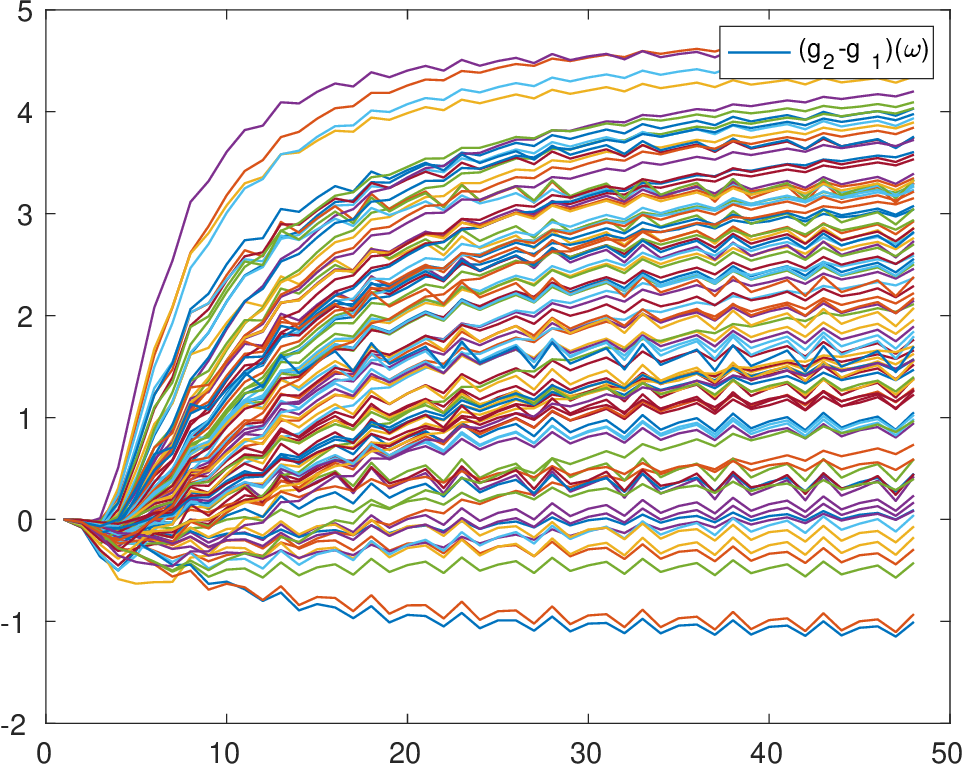}\;
  \includegraphics[width=0.32\textwidth]{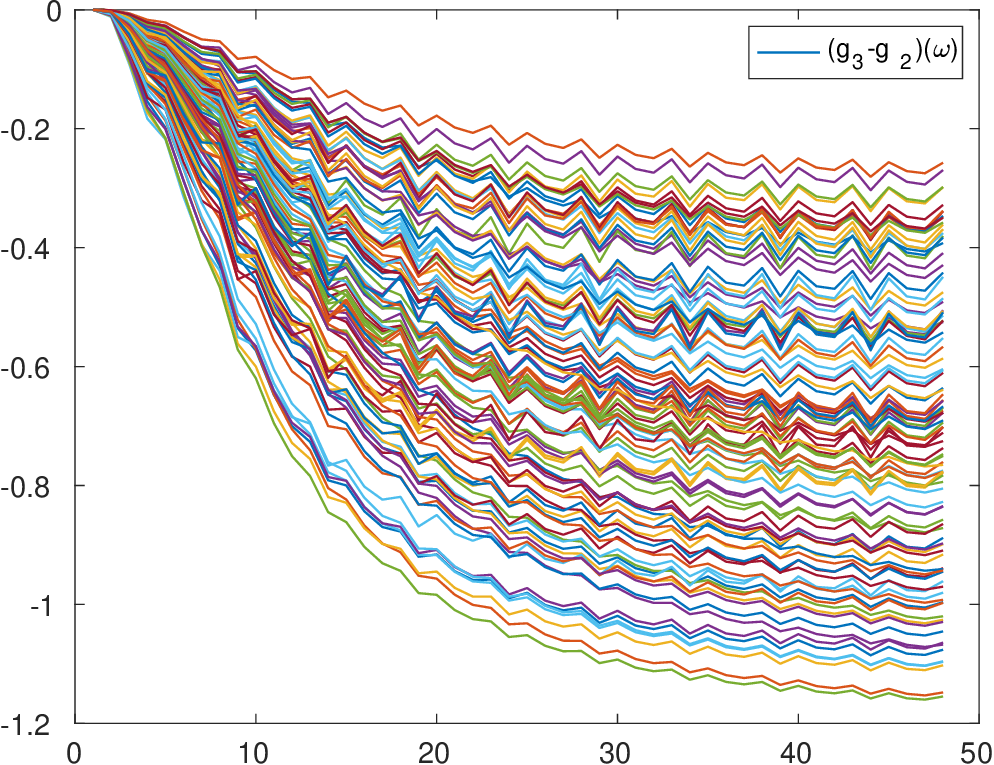}\;
\end{center}
 \caption{100 realisations of $g_1-g_0$ (left), $g_2-g_1$ (center), $g_3-g_2$ (right) {\color{black} vs. time $t$}. Here QoI $g_{\ell}$ is the integral value $Q_s(t)$ as in \eqref{eq:integral_Qs} computed over $\Delta_9$ for $t\in[\tau, 48\tau]$.}
 \label{fig:100realisations_L3579}
 %dima_Henry_points_pdfs_diffs_6Apr.m
\end{figure}

Figure~\ref{fig:100realisations_L3579} shows 100 realisations of $g_1-g_0$ (left), $g_2-g_1$ (center), $g_3-g_2$ (right). Here QoI $g_{\ell}$ is the integral value as in \eqref{eq:integral_Qs} computed  over a subdomain $\Delta_9$ for $t\in[\tau, 48\tau]$. The amplitude of $g_1-g_0$ (on the left) achieves $\approx 19$, of $g_2-g_1$ (in the middle) achieves $\approx 5.5$, and the amplitude of $g_3-g_2$ (on the right) achieves 1.2 (i.e. we observe decay by factor $\approx 4$). 

Table~\ref{tab:adaptiveTS_times} contains average computing times, which are necessary to estimate the number of samples $m_{\ell}$ at each level $\ell$. The fifth column contains the average computing time, and the sixth and seventh columns contain the shortest and longest computing times. The computing time for each simulation varies depending on the number of iterations, which depends on the porosity and permeability. We observed that, after $\approx 6016$~s, the solution is almost unchanging; thus, we perform the experiment only for $t\in [0, T]$, where $T=6016$. For example, if the number of time steps is $r_{\ell}=94$ (Level 0 in Table~\ref{tab:adaptiveTS_times}), then the time step $\tau = \frac{T}{r_{\ell}}=\frac{6016}{94}=64$~s. 
%Table~\ref{tab:fixedTS_times} is for the settings when the time step $\tau$ is fixed and $\tau=1$ sec. (very fine mesh). We observe that the averaged computing time is growing linear with $n_{\ell}$.

% \begin{table}[htbp!]
% \begin{center}
% \begin{tabular}{|l|l|l|l|l|l|}
% \hline 
% \multirow{2}{*}{Level $\ell$}& 
% \multirow{2}{*}{$n_{\ell}$} &
% \multirow{2}{*}{$r_{\ell}$} &
% \multicolumn{3}{|c|}{Computing times}\\
% \cline{4-6}
%     &&                                  & averaged & min. & max.\\
% \hline
% 0 &    1122 & 6016 &    23.7 & 22   & 25   \\ \hline % 512 grid elements
% 1 &    4290 & 6016 &    48.0 & 46   & 50   \\ \hline % 2048
% 2 &   16770 & 6016 &   133.5 & 126  & 142  \\ \hline % 8192
% 3 &   66306 & 6016 &   501.4 & 455  & 514  \\ \hline % 32768
% 4 &  263682 & 6016 &  1927.4 & 1644 & 1960 \\ \hline % 131072
% 5 & 1051650 & 6016 &  8138.0 & 6453 & 8777 \\ \hline % 524288
% \end{tabular}
% \caption{Number of the degrees of freedom $n_{\ell}$, number of time steps $r_{\ell}$, averaged, minimal, maximal computing times on each level $\ell$.} %RunRechargeRanPoro}
% \label{tab:fixedTS_times}
% \end{center}
% \end{table}
%

%Table~\ref{tab:adaptiveTS_times} lists computing times for the settings when 
%The time step $\tau$ is adaptive and changing from $\tau=\frac{6016}{128}=32$~s (very coarse mesh) to $\tau=\frac{6016}{6016}=1$~s (finest mesh). Starting with level $\ell=2$, the average time increases by a factor of eight. These numerical tests confirm the theory in Eq.~(\ref{eq:CompComplexity}), stating that the numerical solver is linear w.r.t. $n_{\ell}$ and $ r_{\ell}$.

%
\begin{table}[htbp!]
\begin{center}
\begin{tabular}{|l|l|l|l|l|l|l|}
\hline 
\multirow{2}{*}{Level $\ell$}& 
\multirow{2}{*}{$n_{\ell}$, ($\frac{n_{\ell}}{n_{\ell-1}}$)} &
\multirow{2}{*}{$r_{\ell}$, ($\frac{r_{\ell}}{r_{\ell-1}}$)} &
\multirow{2}{*}{$\tau_{\ell}=6016/r_{\ell}$} &
\multicolumn{3}{|c|}{Computing times ($s_{\ell}$), ($\frac{s_{\ell}}{s_{\ell-1}}$)}\\
\cline{5-7}
    &&   &                               & average & min. & max.\\
\hline
0 &    153       &   94     & 64 &     0.6      &    0.5 &     0.7 \\ \hline % 128 grid elem.
1 &   2145 (14)  &  376 (4) & 16 &     7.1 (14) &    6.9 &     8.7 \\ \hline % 2048
2 &  33153 (15.5)& 1504 (4) &  4 &   252.9 (36) &  246.2 &   266.2 \\ \hline % 32768
3 & 525825 (15.9)& 6016 (4) &  1 & 11109.8 (44) & 9858.4 & 15506.9 \\ \hline % 524288
\end{tabular}
\caption{Number of degrees of freedom $n_{\ell}$, number of time steps $r_{\ell}$, step size in time $\tau_{\ell}$, average, minimal, and maximal computing times on each level $\ell$.} %PeriodicRanRecharge-RanPoro-600-intervals, pure times - without the output
\label{tab:adaptiveTS_times} % PeriodicRanRecharge-RanPoro-600-intervals.
\end{center}
\end{table}

%
% Figure~\ref{fig:mean_var_diffs_levels} shows the mean (left) and the variance (right) of the solution for $t=1\ldots 48$ computed on meshes $L=\{0,1,2,3,4,5\}$. Particularly, one can see a very small difference between the mean values (on the left) computed on all meshes. The variances are also very similar.

% \begin{figure}[htbp!]
% \begin{center}
%   \includegraphics[width=0.4\textwidth]{figs/concentr_means_L4_9_point5_RanRechargeRanPoro.jpg}\;
%   \includegraphics[width=0.4\textwidth]{figs/concentr_var_L4_9_RanRechargeRanPoro.jpg}
%   \caption{The mean (left) and the variance (right) for $p_3=(1.60, −0.95),$ $\ell=0,\dots,5$, $t=1\ldots 48$.}
%     \label{fig:mean_var_diffs_levels}
% \end{center}  
% %dima_compare_MC_with_MLMC.m
% \end{figure}
%

\begin{figure}[htbp!]
\begin{center}
 \includegraphics[width=0.49\textwidth]{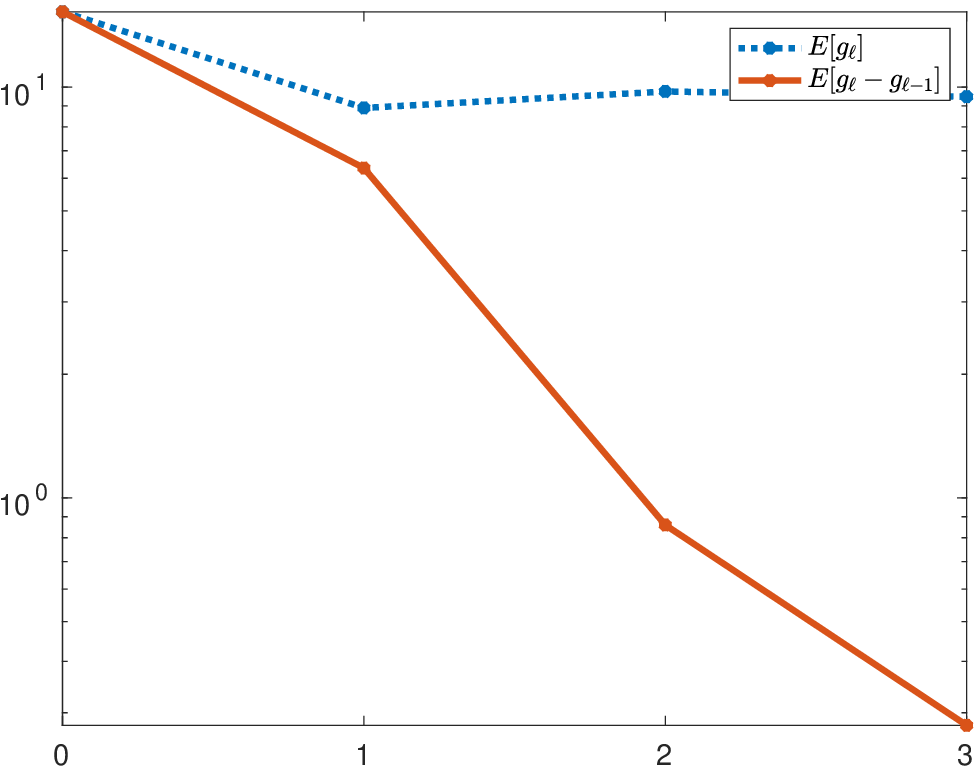}\;
  \includegraphics[width=0.49\textwidth]{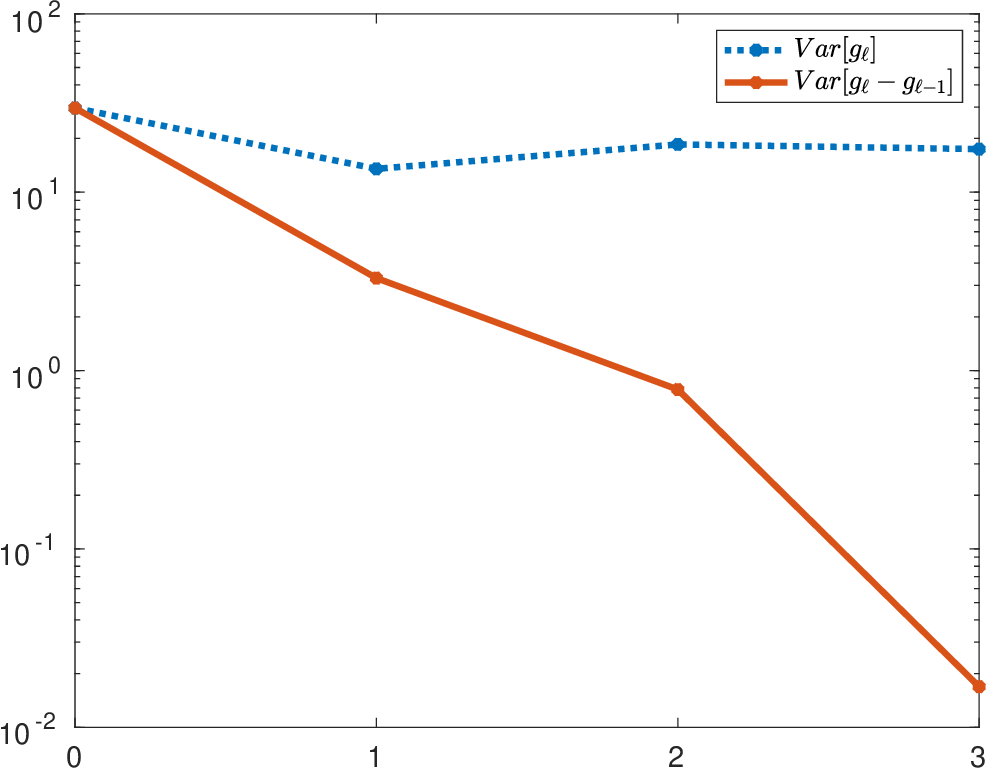}
  \caption{(left) Decay of $\EXP{g_{\ell}-g_{\ell-1}}$ and $V_{\ell}$ in log-scale computed for levels 0,1,2,3 (horizontal axis). The QoI is a subdomain integral $Q_9$ at $t=10\tau$.}
    \label{fig:decayMC_MLMC}
\end{center}  
%/Users/litvinenko/Documents/Henry/MATLABcode/dima_Henry_points_pdfs_diffs_12Mai.m
%OLD dima_compare_MC_with_MLMC.m
\end{figure}

\textbf{Test B2.} {\color{black} The QoI $g_{\ell}$ is the integral $Q_9$ at $t=10\tau$ (see \eqref{eq:integral_box}).} In Figure~\ref{fig:decayMC_MLMC} we compare the decay of $\EXP{g_{\ell}}$ and $\EXP{g_{\ell}-g_{\ell-1}}$ (left) and of $\Var{g_{\ell}}$ and $\Var{g_{\ell}-g_{\ell-1}}$ (right). We can see that the decay of $g_{\ell}-g_{\ell-1}$ is much faster.

\begin{figure}[htbp!]
\begin{center}
 %old \includegraphics[width=0.4\textwidth]{figs/weak_decay_RanRechargeRanPoro.jpg}\;
 %old \includegraphics[width=0.4\textwidth]{figs/strong_decay_RanRechargeRanPoro.jpg}
 \includegraphics[width=0.49\textwidth]{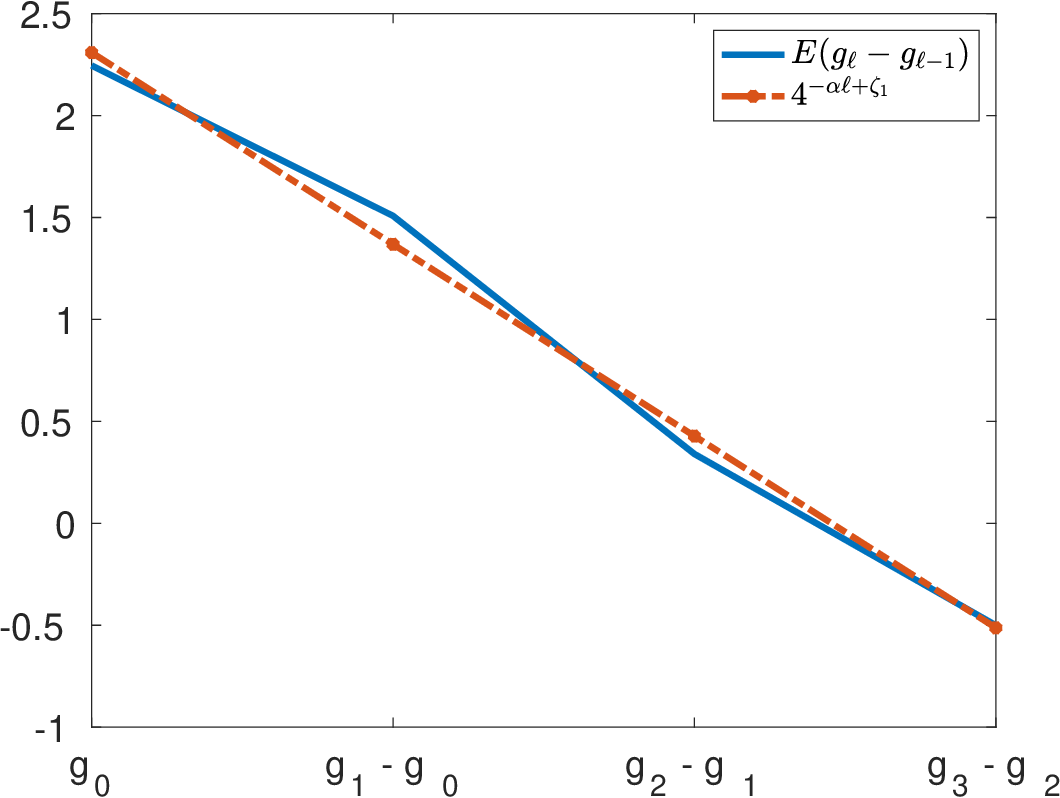}\;
  \includegraphics[width=0.49\textwidth]{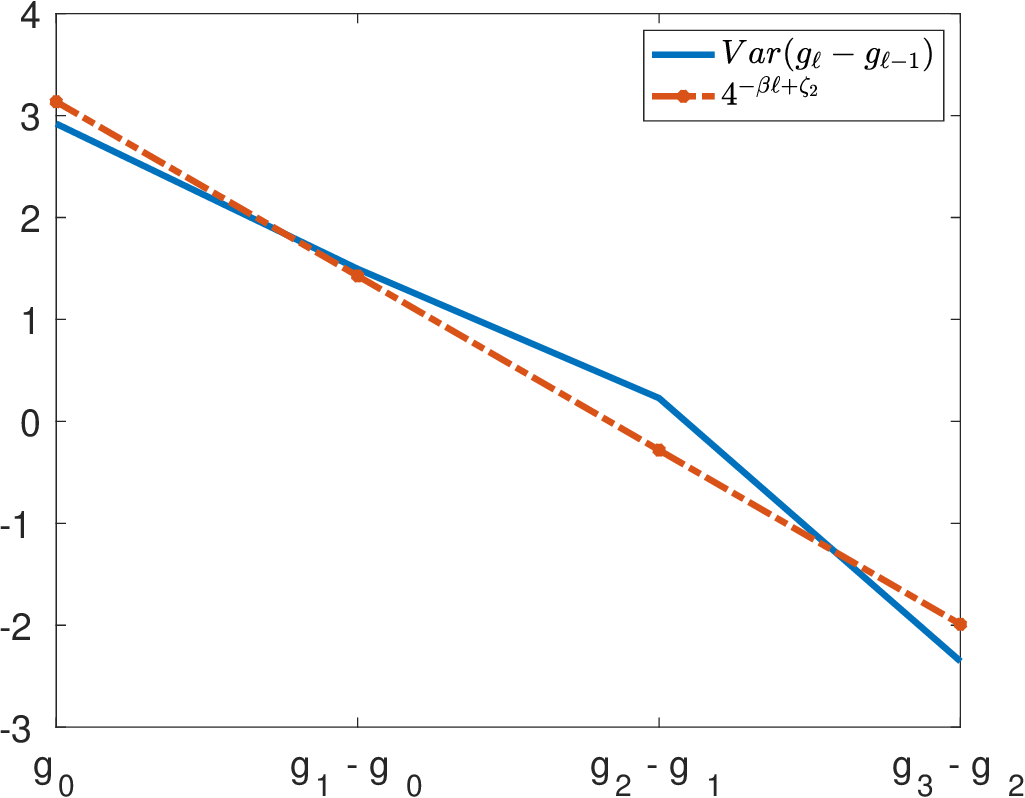}
  \caption{(left) Weak ($\alpha=0.9$, $\zeta_1=3.2$) and (right) strong ($\beta=1.7$, $\zeta_2=4.8$) convergences in log-scale computed for levels 0,1,2,3 (horizontal axis). The QoI is a subdomain integral $Q_9$ at $t=10\tau$.}
    \label{fig:weak_strong}
\end{center}  
%/Users/litvinenko/Documents/Henry/MATLABcode/dima_Henry_points_pdfs_diffs_11Mai.m
%OLD dima_compare_MC_with_MLMC.m
\end{figure}

\textbf{Test B3.} The QoI is a subdomain integral $Q_9$ of $\sol$ over the subdomain $\Delta_9$ - a domain around the point $(x,y)_9 = (1.65, -0.75)$ (see Fig.\ref{fig:Henry2d-scheme} and \eqref{eq:12points}).
The slopes in \eqref{fig:weak_strong} can be used to estimate the rates of weak (shown on the left) and strong (shown on the right) convergences (defined in \eqref{eq:weak_error_model}-\eqref{eq:strong_error_model}). Both graphs are in logarithmic scale. 
The points on the horizontal axis correspond to the QoI calculated at levels $0,1,2$ and $3$.
We fit the graph of $\log_4(\EXP{g_{\ell} - g_{\ell-1}})$ with a linear function $-\alpha\cdot \ell + \zeta_1$. The coefficients obtained are $\alpha\approx 0.9$ and $\zeta_1=3.25$. Furthermore, after fitting the graph of $\log_4(\Var{g_{\ell} - g_{\ell-1}})$ by a linear function $-\beta\cdot \ell + \zeta_2$, we obtained coefficients $\beta\approx 1.7$ and $\zeta_2=4.8$ (see Theorem~\ref{thm:costMLMC}). 
%

% We use computed variances $V_{\ell}$ and computing times (work) $s_{\ell}$ 
% from Table~\ref{tab:adaptiveTS_times} to estimate the optimal number of samples $m_{\ell}$ and compute the telescopic sum from \eqref{eq:A} to approximate the expectation.
%\textcolor{black}{TODO: Alex, add how much we won in time}.

%Table~\ref{tab:M_ell} lists $m_{\ell}$ for a given total variance $\varepsilon^2$:
% \begin{table}[htbp!]
% \begin{center}
% \begin{tabular}{|l|l|l|l|l|l|l|} \hline
% level, $\ell$ & 0 & 1 & 2& 3 & 4 &5 \\ \hline 
% $s_{\ell}$ & 1.156 & 4.113  &20.382 &139.0   &993.0  & 8053.0 \\ 
% $V_{\ell}$ & 1.4e-5& 0.2e-5 & 0.5e-6& 0.1e-6 &0.5e-7 & 1e-7 \\ \hline
% $m_{\ell}(\epsilon^2$ =5e-6$)$ &35  &   7  &   2  &   1  &   1  &   1 \\ \hline
% $m_{\ell}(\epsilon^2$ =1e-6$)$ &172  &  35  &   8   &  2  &   1   &  1 \\ \hline
% $m_{\ell}(\epsilon^2$ =5e-7$)$ & 343  &  69 &   16  &   3  &   1  &   1 \\ \hline
% $m_{\ell}(\epsilon^2$ =1e-7$)$ &
% 1714    &     344    &      78      &    14     &      4       &    2 \\ \hline
% \end{tabular}
% \caption{Number of samples $m_{\ell}$ computed using Eq.~(\ref{eq:M_ell}) as a function of the total variance $\epsilon^2$.}
%   \label{tab:M_ell}
% \end{center}
% %Generated in dima_M_ell.m
% \end{table}
\begin{remark}
Other QoIs have other parameter values. For example, for the freshwater integral $Q_{FW}$ the coefficients are:
$\alpha = 1.82$, $\zeta_1=  1.95$,
$\beta =2.5$,  $\zeta_2= -0.67$. For the salt integral $Q_{S}$ parameters are:
$\alpha = 1.92$, $\zeta_1 = 5.9$,
$\beta = 2.5$, $\zeta_2=8.0$.
\end{remark}
%\textcolor{black}{2Alex: put here profit compare to MC. By the way, can %MC600 achieve such total variance?}

% \begin{figure}[htbp!]
% \centering
% \includegraphics[width=0.49\textwidth]{figs/meanMC200.png}\,
% \caption{(First row)}
% \label{fig:1}
% \end{figure}%
%
%
% Table~\ref{tab:M_ell} lists $m_{\ell}$ for a given total variance $\varepsilon^2$:

% \begin{table}[htbp!]
% \begin{center}
% \begin{tabular}{|l|l|l|l|l|} \hline
% $\varepsilon$ & $m_0$& $m_1$ & $m_2$& $m_3$ \\ \hline
% 0.5    & 1   &         0     &        0     &     0 \\ \hline
% 0.1  & 44  &        5     &       0     &     0 \\ \hline
% 0.05  & 362 &        43    &      3      &     0 \\ \hline
% 0.01 & 16672&        1990   &      120     &     4 \\ \hline
% \end{tabular}
% \caption{The number of samples $m_{\ell}$ computed using Eq.~(\ref{eq:M_ell}) as a function of the total variance $\epsilon^2$.}
%   \label{tab:M_ell}
% \end{center}
% %Generated in /Users/litvinenko/Documents/Dima_Henry/dima_compare_MCwithMLMC_29May.m
% \end{table}
% %

\begin{figure}[htbp!]
\begin{center}
  \includegraphics[width=0.6\textwidth]{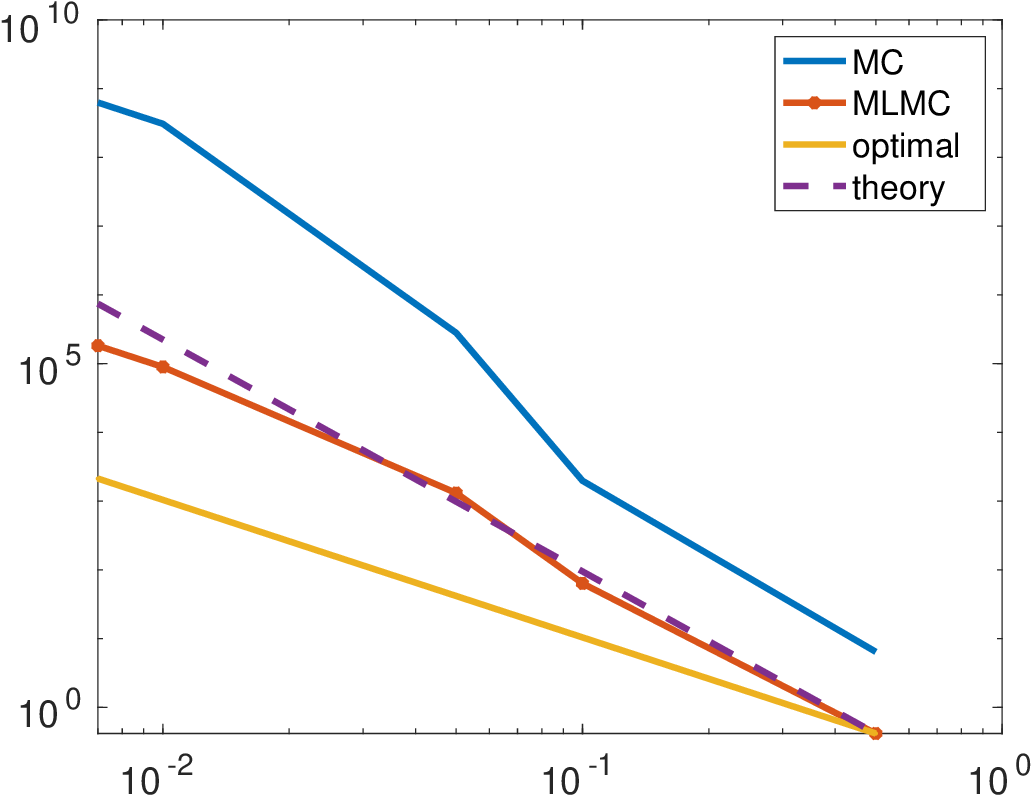}\;
  \caption{Comparison of MC and MLMC for different {\color{black} values of} $\varepsilon$ ($x$-axes).}
    \label{fig:comparison_MC_MLMC_L}
%/Users/litvinenko/Documents/Dima_Henry/dima_compare_MCwithMLMC_29May.m
\end{center}    
\end{figure}

\textbf{Test B4.} Using \eqref{eq:total_cost_MLMC}
we can compute the total cost of MLMC $S=2\cdot \varepsilon^{-2}\left( \sum_{\ell=0}^L \sqrt{V_{\ell} s_{\ell}}\right)^2$, which we compared with the estimated cost of the standard MC method $S_{\mbox{MC}}=2\cdot \varepsilon^{-2} s_L \cdot \Var{g_0}$. It is common in MLMC papers to perform such comparison. The results of this comparison are visible in Table~\ref{tab:comparison_ML_MLMC} and in Fig.~\ref{fig:comparison_MC_MLMC_L}. We can see that MLMC outperforms MC by a factor of $\approx 30 - 3000$ depending on the MSE accuracy $\varepsilon^2$. We can also see the estimated number of required levels $L$ and the numbers of samples $\{m_0,m_1,m_2,m_3\}$ for each accuracy $\varepsilon^2$. These results are also visualized in Fig.~\ref{fig:comparison_MC_MLMC_L}. Additionally, to the MLMC and MC curves, we added two other graphics. The first one (violet dotted line, labelled ``theory'') corresponds to the curve
$\varepsilon^{-(2+\frac{\tilde{d}\cdot \gamma - \beta}{\alpha})}$ (see \eqref{eq:mlmc_iso_work} from Theorem~\ref{thm:costMLMC} for $\beta < \tilde{d}\gamma$, with $\beta\approx 1.7$, $\tilde{d}=3$, $\gamma=1$, $\alpha=0.9$). The second (yellow line) corresponds to
$\varepsilon^{-2}$, and we labelled it ``optimal''. Thus, the MLMC slope is $2+\frac{\tilde {d}\cdot \gamma - \beta}{\alpha}=2+\frac{3\cdot 1 - 1.7}{0.9}=3.44$, and the MC slope $\varepsilon^{-2-3\cdot 1/0.9})=\mathcal{O}(\varepsilon^{-5.3})$. Here $\gamma=1$ and $d=3$ (spatial dimension 2 plus temporal dimension 1), and the required condition $0.9\geq \frac{1}{2}\min(1.7, 3)$ is satisfied. 
\begin{table}[htbp!]
\begin{center}
\begin{tabular}{|l|l|l|l|l|}
\hline 
$\varepsilon$   & 0.1 & 0.05 & 0.01 & 0.007 \\   \hline
$\varepsilon^2$   & 0.01 & 0.0025 & 0.0001 & 0.000049 \\   
\hline
MC cost, $S_{MC}$ & $2.0\cdot 10^3$& $2.8\cdot 10^5$& $3.1\cdot 10^8$ & $6.3\cdot 10^8$\\ \hline
MLMC cost, $S$ &  $6.4\cdot 10^1$ &  $1.3 \cdot 10^3$ & $8.9\cdot 10^4$ & $1.8\cdot 10^5$\\ \hline
required $L$ & 2 & 3 & 4 & 4 \\ \hline
 $\{m_0,m_1,m_2,m_3\}$ & $\{44, 5,0,0\}$& $\{404, 49,3,0\}$& $\{16672,   1990,   120,   4\}$& $\{34024, 4062, 245, 7\}$ \\ \hline
\end{tabular}
\caption{Comparison of MC and MLMC and the number of samples on each level vs $\varepsilon$.}
\label{tab:comparison_ML_MLMC}
\end{center}
\end{table}

%
%\newpage
\section{Conclusion}
\label{sec:Conclusion}
{\color{black}
We investigated the applicability and efficiency of the MLMC approach to the Henry-like problem with uncertain porosity, permeability and recharge. These uncertain parameters were modelled by random fields with three independent random variables. Permeability is a function of porosity. Both functions are time-dependent, have multi-scale behaviour and are defined for two layers. The numerical solution for each random realisation was obtained using the well-known ug4 parallel multigrid solver. The number of random samples required at each level was estimated by calculating the decay of the variances and the computational cost for each level. These estimates depend on the minimisation function defined in \eqref{eq:goal_function}.
}

{\color{black}
The MLMC method was used to compute the expected value and variance of several QoIs, such as the solution at a few preselected points $(t,\bx)$, the solution integrated over a small subdomain, and the time evolution of the freshwater integral. We have found that some QoIs require only 2-3 mesh levels and samples from finer meshes would not significantly improve the result. Other QoIs require more grid levels. Note that a different type of porosity in \eqref{eq:poro_2levels} may lead to a different conclusion.
}

The numerical results confirm that the computational cost of the MLMC method is lower than that of the MC method. Therefore, sampling at different mesh levels is useful and helps to reduce the overall computational cost.\\
\textbf{Constraints.} 1. {\color{black} It may happen that the QoIs calculated at different mesh levels are very similar. In this case the default (Q)MC is sufficient. 2. The time dependence is challenging: the optimal number of MLMC samples depends on the QoI. For example, this number may be large for the solution computed at one point $(t,\bx)$ and small for the solution computed at another time. 3. Each new QoI requires new estimates of all parameters $\alpha$, $\beta$, $\gamma$, and the resulting MLMC graphs may be very different.}\\
\textbf{Future work.} 1. It would be beneficial to consider a more complicated/multiscale/realistic porosity and permeability with more random variables. 2. A more advanced version of MLMC may give better estimates of the number of levels $L$ and the number of samples on each level $m_{\ell}$. 3.
Known experimental and measured data of porosity, permeability, velocity or mass fraction could be used to identify unknown parameters and minimise uncertainties \cite{Litv_HLIBCov2020,LitvGenton19,Litv_params2016,Rosic2013}.
\section*{Acknowledgments}
For computing time, this research used Shaheen II, which is managed by the Supercomputing Core Laboratory at the King Abdullah University of Science and Technology (KAUST) in Thuwal, Saudi Arabia. We thank the KAUST HPC support team for their assistance with Shaheen II. This work was supported by the Alexander von Humboldt Foundation. The authors also appreciate the comments and remarks of the reviewers, which helped us to improve this work.

\bibliographystyle{siam}
% WHERE IS THIS FILE ? \bibliographystyle{spmpsci}
%\bibliography{new_article_Sydney_about_sampling, matthies_BU_paper-1, mybib, references}
%\bibliography{litvinenko_dolgov_khoromskij}
%\bibliography{MoCa}

\begin{table}[htbp!]
\begin{tabular}{|c|l|}
\hline
\multicolumn{2}{|c|}{\textbf{Notation}} \\ \hline
%${\Nset}$        &Natural numbers \\ \hline
RV & \textcolor{black}{random variable} \\ \hline
QoI $g$ & quantity of interest $g$ \\ \hline
(q)MC & (quasi-) Monte Carlo \\ \hline
MLMC & Multilevel Monte Carlo \\ \hline
$L$ & number of mesh levels \\ \hline
$\mathcal{D}$ & computational spatial domain \\ \hline
$\mathcal{D}_0,\mathcal{D}_1,\ldots,\mathcal{D}_L$ & hierarchy of spatial meshes \\ \hline
$\mathcal{T}_0,\mathcal{T}_1,\ldots,\mathcal{T}_L$ & hierarchy of temporal meshes \\ \hline
$s_{\ell}$, $S$ & complexity on level $\ell$, total complexity\\ \hline
$h_{\ell}$ (or $h$), $n_{\ell}$ & spatial step size and number of spatial degrees of freedom on level $\ell$ \\ \hline
$\tau_{\ell}$ (or $\tau$), $r_{\ell}$ & time step size and number of time steps on level $\ell$ \\ \hline
$m_{\ell}$ & number of samples (scenarios) on level $\ell$ \\ \hline
$\omega$, $\xib(\omega)$ & random event and random vector \\ \hline
$\EXP{\cdot}$, $\Var{\cdot}$           & expectation and variance w.r.t. RV $\xib$\\ \hline
$\EXPt{\cdot}$         &  expectation w.r.t. the time $t$\\ \hline
$\vTheta$ & multidimensional domain of integration in parametric space \\ \hline
$\poro(\bx,\omega)$ & porosity random field  \\ \hline
$\perm(\bx,\omega)$ & permeability random field \\ \hline
$\dens(\bx, \omega)$ & density random field  \\ \hline
$\dvel(t,\bx, \omega)$ & volumetric velocity \\ \hline
%$\conc(t,x,\omega)$ & mass fraction\\ \hline
$\disp$ &  tensor field $\disp = \disp (\dvel)$: molecular diffusion and dispersion of salt \\ \hline
$\bar{\kappa}(\bx)$ & expectation of $\kappa(\bx,\omega)$ \\ \hline
$d$ & physical (spatial) dimension \\ \hline
$\hat{d}$ & total dimension, $\hat{d} = d+1$ \\ \hline
% $\param$ & high-dimensional parameter \\ \hline
%$\f$            & the right-hand side \\ \hline
$\sol=\sol(t,\bx,\omega)$          & mass fraction of salt (solution of the problem)\\ \hline
\end{tabular}
\label{tab:notations}
\end{table}

\end{document}